\documentclass[aps,pra,letterpaper,twocolumn,showpacs,amssymb]{revtex4-1}

\usepackage{amsmath}
\usepackage{graphicx}

\newcommand{\w}{\omega}
\newcommand{\p}{\prime}
\newcommand{\overbar}[1]{\mkern 1.5mu\overline{\mkern-1.5mu#1\mkern-1.5mu}\mkern 1.5mu}

\begin{document}

\title{Narrowband biphoton generation in the group delay regime}

\author{Luwei Zhao}
\author{Yumian Su}
\email{Electronic address: yumiansu@ust.hk}
\author{Shengwang Du}
\email{Electronic address: dusw@ust.hk}
\affiliation{Department of Physics, The Hong Kong University of
Science and Technology, Clear Water Bay, Kowloon, Hong Kong, China}

\date{\today}% It is always \today, today,
             %  but any date may be explicitly specified
\begin{abstract}

 We study narrow-band biphoton generation from spontaneous four-wave mixing with electromagnetically induced transparency in a laser cooled atomic ensemble. We compare two formalisms in the interaction and Heisenberg pictures, and find that they agree in the low gain regime but disagree in the high gain regime. We extend both formalisms accounting the non-uniformity in atomic density and the driving laser fields. We find that for a fixed optical depth and a weak and far-detuned pump laser beam, the two-photon waveform is independent of the atomic density distribution. However, the spatial profiles of the two driving laser beams have significant effects on the biphoton temporal waveform. We predict that waveform shaping in time domain can be achieved by controlling the spatial profiles of the driving laser fields.
\end{abstract}

%\pacs{42.50.Dv, 42.65.Lm, 42.50.Gy}

\maketitle

\section{Introduction} \label{sec:Introduction}

Entangled photon pairs, termed biphotons, have been studied extensively for a range of quantum applications, including quantum information processing, quantum communication and quantum cryptography~\cite{GisinRMP2002,BraunsteinRMP2005}. To improve its spectral brightness, a lot of efforts have been made to generate narrowband, long coherent biphotons with various methods, e.g., cavity-assisted spontaneous parametric down conversion (SPDC) in nonlinear crystals \cite{OuPRL1999,KuklewiczPRL2006, PanPRL2008, ChuuAPL2012, ChuuPRA2011} and spontaneous four-wave mixing (SFWM) in atomic systems \cite{BalicPRL2005, DuPRL2007, DuPRL2008, SrivathsanPRL2013, ZhaoOPTICA2014}. These biphotons can be used to produce narrow-band heralded single photons as the time origin is established by one of the paired photons. With the long coherence time ranging from several hundred nanoseconds to microseconds, the heralded single-photon waveform can be shaped by an electro-optical modulator \cite{SinglePhotonEOM}. Their capability to interact with atoms resonantly has been applied for observing single-photon optical precursor\cite{SinglePhotonPrecursor}, improving the storage efficiency of optical quantum memory \cite{ZhouOE2012}, and coherently controlling absorption and reemission of single photons in two-level atoms~\cite{ZhangPRL2012}. Other applications include single-photon differential-phase-shift quantum key distribution~\cite{LiuOE2013}. However, external amplitude modulation causes unavoidable loss to the single photons and also introduces noise. The ideal way to create a desired biphoton waveform is to start from biphoton generation, i.e., to control the driving field and the medium. For the broadband entangled photons generated by SPDC, Valencia \textit{et al.}~\cite{ValenciaPRL2007} demonstrated shaping the joint spectrum by controlling the spatial shape of the pump beam, which is the first spatial-to-spectrum mapping on biphotons. For SFWM narrowband biphotons source with electromagnetically induced transparency (EIT), the two-photon correlation function can be shaped by periodically modulating the classical driving fields in time domain \cite{DuPRA2009, ChenPRL2010}.

In the literature, there are two formalisms to model the EIT-assisted SFWM process. One is to use the perturbation theory in the interaction picture, in which the interaction Hamiltonian describes the four-wave mixing process and determines the evolution of the two-photon state vector \cite{Wen2006, Wen2007_1, Wen2007_2, Wen2008, Du}. This gives a clear picture of the biphoton generation mechanism. The other is developed in the Heisenberg picture~\cite{Kolchin, Ooi, WenRubin, Braje} with the evolution of field operators. In all these previous theoretical models, the atomic density and driving field amplitude are spatially uniform and thus the effect of their non-uniformity has not been studied.

Unlike the rectangle-shaped biphoton waveform predicted previously in the group delay regime \cite{BalicPRL2005, Du, Kolchin}, in our recent experiment \cite{ZhaoOPTICA2014} we produced a Gaussian-like biphoton waveform at high atomic optical depth (OD). This shape cannot be explained by current theoretical models, and this is one of our motivations to extend the existing models.

In this paper, we explore the following points: (1) \textit{Comparison of both models}. We compare the two formalisms in the interaction and Heisenberg pictures and show that in low parametric gain regime both agree well. (2) \textit{Non-uniformity}. We extend the existing theories by taking into account the non-uniformity in the atom distribution, the pump, and the coupling laser intensity distribution in the longitudinal direction of the atomic cloud. we show that the profiles of the pump and coupling laser intensities have significant effects on the biphoton waveform. (3) \textit{Waveform shaping}. By controlling the spatial profile of the driving field, one can shape the biphoton waveform in time domain. On the other hand, the time-domain waveform of the photon pairs allows us to retrieve information on the spatial profile of the pump and coupling laser beams.

The paper is organized as follows: In Sec.~\ref{theory}, we describe the biphoton generation in two approaches: (1) state vector theory in the interaction picture and (2) coupled operator equations in the Heisenberg picture. Sec.~\ref{Numerical_result} gives the numerical results of the models. We first analyze the photon properties, then show that the two approaches are equivalent in the low parametric gain regime. We then propose to shape and engineer biphoton temporal waveform with various spatial profiles of the driving lasers in Sec.~\ref{Discussion}. We give our conclusions in Sec.~\ref{conclusion}.
%-------------------------------------------------------------------------------------------------------------------------

%--------------------------------------------------------------------------------------------------------------------------

\section{Theoretical Framework} \label{theory}

In this paper, we study the EIT-assisted SFWM biphoton generation in both the interaction and Heisenberg pictures. Extending from previous models, we take into account the non-uniformity of atomic density and the spatial profiles of the driving fields. Although both pictures are equivalent, some physics insights are clearer in one picture or the other. In the interaction picture, using perturbation theory, the evolution of the photon state describes more clearly how the biphotons are generated, but the system loss and gain can not be fully accounted. On the other side, the Heisenberg formalism provides a more accurate calculation of the experimental coincidence counts including multi-photon events and accidental coincidences, but the two-photon state is not clearly resolved. We compare both models by exploring the scenario where the atomic density, the intensities of the pump and coupling lasers are not uniform along the length of the atomic cloud.

\begin{figure}[h]
%\vskip -1em
\centering
\includegraphics[width=\linewidth]{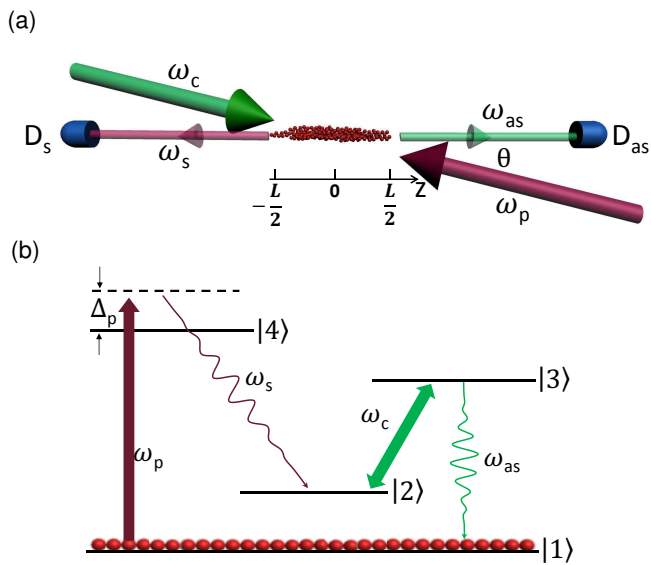}
%\vspace{-1em}
\caption{(Color online) A schematic of four-wave mixing to generate biphotons. (a) Optical configuration. (b) Atomic energy level diagram. The pump laser couples the transition $|1\rangle \to |4\rangle$ with a detuning $\Delta_p$, while the coupling laser is on resonance with the transition $|2\rangle\rightarrow|3\rangle$. Paired Stokes and anti-Stokes photons are spontaneously produced following the SFWM path.}
\label{schematic}
\end{figure}

Figure \ref{schematic} is a schematic of biphoton generation from a four-level double-$\Lambda$ cold atomic medium with a length L. The atoms are identical and prepared in the ground state $|1\rangle$. A pump laser with frequency $\omega_p$ excites the transition $|1\rangle\rightarrow|4\rangle$ with a detuning $\Delta_p$, and a coupling laser on resonance with the transition $|2\rangle\rightarrow3\rangle$ propagates in the opposite direction of the pump laser. Phase-matched, counter-propagating Stokes ($\omega_s$) and anti-Stokes ($\omega_{as}$) photon pairs are spontaneously generated following the SFWM path. The coupling laser renders the EIT window for the resonant anti-Stokes photons that travel with a slow group velocity \cite{EIT, EIT02}. The Stokes photons travel in the atomic cloud nearly with the speed of light in vacuum and a negligible Raman gain. The counter-propagating pump-coupling beams are aligned with a small angle $\theta$ with respect to the biphoton generation longitudinal $z$-axis. Assuming that the pump laser is weak and far-detuned from $|1\rangle \to |4\rangle$ transition, the majority of the atoms are in the ground state $|1\rangle$.

\subsection{Interaction Picture}

Here we study the biphoton generation in the interaction picture with a focus on the evolution of the two-photon state. We extend the previous theory by Du \textit{et al.} \cite{Du} to take into account the spatial non-uniformity of the nonlinear interaction. With $z$ direction being the longitudinal direction of the biphoton generation as shown in Fig.~\ref{schematic}, the electric field in this direction is given by $E(z,t) = [E^{(+)}(z,t) + E^{(-)}(z,t)]/2$, where $E^{(\pm)}$ are positive/negative frequency parts. Assuming that the counter-propagating pump and coupling laser beams are classical fields and undepleted in the atomic medium, and their projections on the longitudinal $z$-axis are described as
\begin{equation}\label{efield_pc}
\begin{split}
E_p^{(+)}(z,t) = & E_p(z) e^{i(-k_p z \cos\theta- \w_p t)} \\
E_c^{(+)}(z,t) = & E_c(z) e^{i(k_c z \cos\theta- \w_c t)},
\end{split}
\end{equation}
where $k_p (k_c)$ is the wavenumber of the pump (coupling) laser field. We treat the single-transverse mode Stokes and anti-Stokes fields as quantized operators
\begin{equation}\label{efield_sas}
\begin{split}
& \hat{E}_s^{(+)}(z,t) =  \sqrt{\frac{2\hbar \w_{s0}}{c\varepsilon_0 A}} \hat{a}_s(z,t)   \\
& \ \ \ \        = \frac{1}{\sqrt{2\pi}} \sqrt{\frac{2\hbar \w_{s0}}{c\varepsilon_0 A}}
							          \int{d\w_s\hat{a}_s(\w_s) e^{i\left[-\int_0^z{k_s(z^\p) dz^\p} - \w_s t\right]} } \\
& \hat{E}_{as}^{(+)}(z,t) =  \sqrt{\frac{2\hbar \w_{as0}}{c\varepsilon_0 A}}\hat{a}_{as}(z,t)  \\
& \ \ \ \           =  \frac{1}{\sqrt{2\pi}}\sqrt{\frac{2\hbar \w_{as0}}{c\varepsilon_0 A}}
									      \int{d\w_{as}\hat{a}_{as}(\w_{as}) e^{i \left[\int_0^z{k_{as}(z^\p) dz^\p} - \w_{as} t\right] }}.
\end{split}
\end{equation}
Here $c$ is the speed of light in vacuum, $\varepsilon_0$ is the vacuum permittivity, $A$ is the single mode cross-sectional area, and $\w_{as0} (\w_{s0})$ is the central frequency of the anti-Stokes (Stokes) photon. $k_s(z^\p)$ and $k_{as}(z^\p)$ are the wavevectors of the Stokes and anti-Stokes fields inside the atomic medium, respectively. $\hat{a}_s(\w_s)$ [$\hat{a}_{as}(\w_{as})$] annihilates a Stokes (anti-Stokes) photon of frequency $\w_s$ ($\w_{as}$), and they satisfy the commutation relation
\begin{equation}\label{commutation_w}
[\hat{a}_s(\w), \hat{a}^\dag_s(\w^\p)] = [\hat{a}_{as}(\w), \hat{a}^\dag_{as}(\w^\p)] = \delta(\w - \w^\p).
\end{equation}

The interaction Hamiltonian that describes the SFWM process is
\begin{equation}\label{H_I}
\begin{split}
& \hat{H}_I = \frac{\varepsilon_0 A}{4}\int_{-\frac{L}{2}}^{\frac{L}{2}} dz\ \chi^{(3)}(z)
       \Big[ E_p^{(+)}(z,t)E_c^{(+)}(z,t)  \\
			& \hspace{8em} \times  \hat{E}_{as}^{(-)}(z,t)\hat{E}_s^{(-)}(z,t) \Big] + \textrm{H.c.},
\end{split}
\end{equation}
where $\chi^{(3)}(z)$ is the third-order nonlinear susceptibility, and is given in Eq.~\eqref{chi_as_3rd_def} in Sec.~\ref{Numerical_result}. $E_s^{(-)}(z,t)$ and $E_{as}^{(-)}(z,t)$ are the Hermitian conjugates of the Stokes and anti-Stokes fields [Eq.~\eqref{efield_sas}], respectively. Substituting the electric fields in Eqs.~\eqref{efield_pc} and \eqref{efield_sas} into Eq.~\eqref{H_I} gives
\begin{equation}\label{H_I_1}
\begin{split}
& \hat{H}_I =   \frac{\hbar\sqrt{ \w_{s0}\w_{as0} } } {4\pi c}
         \int
		          d\w_s dw_{as}
              \int_{-\frac{L}{2}}^{\frac{L}{2}}							
				             dz\ \chi^{(3)}(z) E_p(z)E_c(z)  \\
&   \times                  e^{ -i\int_0^z{\Delta k(z^\p)dz^\p} }
								  \hat{a}_{as}^\dag(\w_{as}) \hat{a}_s^\dag(\w_s)  e^{ -i(\w_p + \w_c - \w_s - \w_{as}) t }  + H.c.,
\end{split}
\end{equation}
where
\begin{equation}
\Delta k(z^\p) \equiv k_{as}(z^\p) - k_s(z^\p) - (k_c - k_p)\cos\theta,
\end{equation}
is the phase mismatching in the atomic cloud.

The two-photon state $|\Psi\rangle$ can be computed in the first-order perturbation theory as
\begin{equation} \label{psi}
\begin{split}
|\Psi\rangle = -\frac{i}{\hbar} \int_{-\infty}^\infty \hat{H}_I |0\rangle \ dt.
\end{split}
\end{equation}
Substituting Eq.~\eqref{H_I_1} into Eq.~\eqref{psi} and integrating over $t$ gives
\begin{equation}\label{psi_1}
\begin{split}
& |\Psi\rangle =  \frac{\sqrt{ \w_{s0} \w_{as0} } }{ 2ic }
                 \int d\w_{as} \int_{-\frac{L}{2}}^{\frac{L}{2}} dz
							   \Big[\chi^{(3)}(z) E_p(z) E_c(z) \\
& \ \ \ \times e^{-i \int_0^z \Delta k(z^\p) dz^\p }
							   \hat{a}_{as}^\dag(\w_{as}) \hat{a}_s^\dag(\w_p + \w_c - \w_{as}) |0\rangle \Big] .
\end{split}
\end{equation}
Note that from Eq.~\eqref{psi} to \eqref{psi_1}, we have made use of the time integral, $\int_{-\infty}^\infty e^{ -i(\w_p + \w_c - \w_s - \w_{as}) t}  dt
                              = 2\pi \delta(\w_p + \w_c -\w_s - \w_{as}) $, which expresses energy conservation of the four-wave mixing process and leads to the frequency entanglement of the two-photon state.

In our setup in Fig.~\ref{schematic}, we neglect the free space propagation (which only cause time shift in measurement) and place the Stokes photon detector $D_s$ at $z= -L/2$ and anti-Stokes photon detector $D_{as}$ at $z=L/2$. The annihilation operators at these two boundaries are
\begin{equation} \label{surface_a}
\begin{split}
\hat{a}_s(t) = & \frac{1}{\sqrt{2\pi}} \int d\w \hat{a}_s(\w)
                   e^{i\int_0^{\frac{L}{2}}k_s(-z^\p)dz^\p -i\w t}, \\
\hat{a}_{as}(t) = & \frac{1}{\sqrt{2\pi}} \int d\w \hat{a}_{as}(\w)
                  e^{i\int_0^{\frac{L}{2}}k_{as}(z^\p)dz^\p -i\w t}
\end{split}
\end{equation}

The two-photon Glauber correlation function can be computed from
\begin{equation} \label{G2_0}
\begin{split}
G^{(2)}(t_s, t_{as}) = & \langle \Psi| \hat{a}_{as}^\dag(t_{as}) \hat{a}_{s}^\dag(t_s)
                                     \hat{a}_{s}(t_{s}) \hat{a}_{as}(t_{as})
												  |\Psi \rangle   \\
								%			& + \langle \Psi| \hat{a}_{as}^\dag(t_{as}) \hat{a}_{as}(t_{as}) |\Psi\rangle
								%			    \langle \Psi| \hat{a}_{s}^\dag(t_{s}) \hat{a}_{s}(t_{s}) |\Psi\rangle \\
									   = & |\langle 0 | \hat{a}_{s}(t_{s}) \hat{a}_{as}(t_{as}) |\Psi \rangle|^2 \\
										 \equiv & | \Psi(t_s, t_{as}) |^2,																						
\end{split}
\end{equation}
where $\Psi(t_s, t_{as})$ is the Stokes--anti-Stokes biphoton amplitude. Substituting Eqs.~\eqref{psi_1} and \eqref{surface_a} and  into Eq.~\eqref{G2_0}, the biphoton amplitude becomes
\begin{equation}\label{biphoton}
\begin{split}
 & \Psi(t_s, t_{as}) =  \frac{\sqrt{ \w_{s0}\w_{as0} } } {i4\pi c} \int d\w_{as} d\w_s^\p d\w_{as}^\p
                        \int_{-\frac{L}{2}}^\frac{L}{2} dz    \\
 &                      \times\Big\{\chi^{(3)}(z)E_p(z)E_c(z) e^{-i\int_0^z \Delta k(z^\p)dz^\p} \\
 &    \times             e^{i\int_0^{\frac{L}{2}}\left[k_s(-z^\p)+k_{as}(z^\p)\right]dz^\p}
  										  e^{-i\w_s^\p t_s - i\w_{as}^\p t_{as} }    \\
 &  \times               \langle 0| \hat{a}_s(\w_s^\p) \hat{a}_{as}(\w_{as}^\p)
                                    \hat{a}_{as}^\dag(\w_{as}) \hat{a}_s^\dag(\w_p+\w_c-\w_{as})
										      |0 \rangle \Big\}
\end{split}
\end{equation}
Using the commutation relation of Eq.~\eqref{commutation_w}, we have $ \langle 0| \hat{a}_s(\w_s^\p) \hat{a}_{as}(\w_{as}^\p)\hat{a}_{as}^\dag (\w_{as})\hat{a}_s^\dag(\w_p+\w_c-\w_{as}) |0\rangle=\delta(\w_s^\p -\w_p - \w_c + \w_{as})\delta(\w_{as}^\p - \w_{as})$. Integrating over $d\w_s^\p$ and $d \w_{as}^\p$, the biphoton amplitude is now
\begin{equation}\label{biphoton_1}
\begin{split}
 & \Psi(t_s, t_{as}) =  \frac{\sqrt{ \w_{s0}\w_{as0} } } {i4\pi c} \int d\w_{as}
                        \int_{-\frac{L}{2}}^\frac{L}{2} dz    \\
 &                      \times\Big\{\chi^{(3)}(z)E_p(z)E_c(z) e^{-i\int_0^z \Delta k(z^\p)dz^\p} \\
 &  \times                e^{i\int_0^{\frac{L}{2}}\left[k_s(-z^\p)+k_{as}(z^\p)\right]dz^\p}
  										  e^{-i(\w_p + \w_c) t_s - i\w_{as} \tau } \Big\} ,
\end{split}
\end{equation}
where $\tau \equiv t_{as} - t_s$.

The wavenumbers of the Stokes and anti-Stokes photons can be described as follows:
\begin{equation}\label{ks_kas}
\begin{split}
k_s(z) = & \frac{\w_{s0}}{c} \sqrt{1+\chi_s(z,\omega_{s0}-\varpi)} \\
k_{as}(z) = & \frac{\w_{as0}}{c} \sqrt{1+\chi_{as}(z,\omega_{as0}+\varpi)},
\end{split}
\end{equation}
where $\omega_{s0}$ ($\omega_{as0}$) is the Stokes (anti-Stokes) photon central frequency, and $-\varpi$ ($\varpi)$ is the Stokes (anti-Stokes) frequency detuning. $\chi_s(z, \omega_{s0}-\varpi)$ and $\chi_{as}(z, \omega_{as0}+\varpi)$, given in Sec.~\ref{Numerical_result} (Eqs.~\eqref{chi_s_def} and \eqref{chi_as_def} ), are the linear susceptibilities of the Stokes and anti-Stokes fields, respectively. Note that $k_s$ and $k_{as}$ are also functions of $\varpi$ through $\chi_s(z, \omega_{s0}-\varpi)$ and $\chi_{as}(z, \omega_{as0}+\varpi)$. Making use of $\omega_p+\omega_c=\omega_{s0}+\omega_{as0}$, we can rewrite Eq.~\eqref{biphoton_1}  as
\begin{equation}\label{biphoton_2}
\Psi(t_s, t_{as}) = \psi(\tau)e^{-i(\w_{as0}t_{as} + \w_{s0}t_s)},
\end{equation}
where the biphoton relative wave amplitude is given by
\begin{equation}\label{biphotonRelative}
\psi(\tau) \equiv  \frac{\sqrt{ \w_{s0}\w_{as0} } } {i4\pi c}  \int d\varpi F(\varpi) Q(\varpi) e^{- i\varpi \tau}.
\end{equation}
Here
\begin{equation} \label{F_deltaomega}
F(\varpi) \equiv \int_{-\frac{L}{2}}^\frac{L}{2} dz
                             \chi^{(3)}(z)E_p(z)E_c(z) e^{-i\int_0^z \Delta k(z^\p)dz^\p},
\end{equation}
\begin{equation}\label{Q_omega}
 Q(\varpi) = e^{i\int_0^{\frac{L}{2}}\left[k_s(-z^\p)+k_{as}(z^\p)\right]dz^\p}.
\end{equation}

%------------------------------------------------------------------------
\subsubsection{No $z$ dependence}
%------------------------------------------------------------------------

If the atomic cloud is homogenous, the pump and coupling lasers have uniform electric fields in the atomic cloud, Eq.~\eqref{F_deltaomega} can be computed analytically and the result is
\begin{equation}
F(\varpi) = L \chi^{(3)}(\varpi) E_p E_c \textrm{sinc}\left(\frac{\Delta k L}{2}\right),
\end{equation}
where, $\Delta k = k_{as} - k_{s} - (k_c - k_p) \cos\theta$, is a function of $\varpi$ only. The biphoton relative wavefunction $\psi(\tau)$ is now
\begin{equation} \label{psi_no_z}
\begin{split}
& \psi(\tau) = \frac{\sqrt{ \w_{s0}\w_{as0} } } {i4\pi c} L E_p E_c \\
& \times\int d\varpi\ \chi^{(3)}(\varpi)\ \textrm{sinc}\left(\frac{\Delta k L}{2}\right)
												e^{i(k_s + k_{as})L/2 }  e^{- i\varpi \tau}.
\end{split}
\end{equation}
 Here $k_s$ and $k_{as}$ are given by Eq.~\eqref{ks_kas}. As the first-order susceptibilities $\chi_s$ and $\chi_{as}$ have no $z$-dependence,
$k_s$ and $k_{as}$ are functions of $\varpi$ only, independent of $z$. The expression of the biphoton amplitude agrees with the previous work by Du \textit{et al.}~\cite{Du}.

For a weak pump laser that is far-detuned, $\chi_s\approx 0$ and $\Delta k$ depends on the group velocity of anti-Stokes photons $V_g$ as $\Delta k\approx \varpi/V_g$. In this case, Eq.~\eqref{psi_no_z} is proportional to the rectangular function $\Pi(\tau;0,L/V_g)$ which ranges from $\tau = 0$ to $L/V_g$~\cite{Du}. Physically, this rectangular waveform can be explained as follows~\cite{Du,Rubin}: the Stokes and anti-Stokes photons are always produced in pairs by the same atom in the atomic cloud. In our experimental setup as illustrated in Fig.~\ref{schematic}, when the photon pairs are produced at the surface $z=L/2$, anti-Stokes photon does not need to go through the atomic medium to arrive at the anti-Stokes detector $D_{as}$. For Stokes photon, as $\chi_s\approx 0$ and $k_s$ is approximately the wavevector in vacuum. Therefore, the Stokes photon travels in the atomic medium in nearly the same velocity as in vacuum. As the length of the atomic medium is very short, the travel time is negligible to reach detector $D_s$. As such, both Stokes and anti-Stokes detectors register a photon almost simultaneously. When the photon pairs are produced at the other surface $z=-L/2$, the anti-Stokes photon has to go through the medium with group velocity $V_g$ to reach detector $D_{as}$, while the Stokes photon does not have to travel through the medium in order to reach detector $D_s$. After $D_s$ registers the Stokes photon, a delay $\tau_g = L/V_g$ later, the anti-Stokes photon reaches detector $D_{as}$. When photon pairs are produced between $-L/2$ to $L/2$, the delay time is between 0 to $\tau_g$. As every atom in the cloud has the same third-order susceptibility and experiences the same pump and coupling laser fields, the probability for producing the photon pairs in every point in the atomic medium is the same. This results in the rectangular waveform.

%-------------------------------------------------------------------------------
\subsubsection{Atomic density is not uniform}

Let's consider the case when the atoms are not distributed uniformly along the longitudinal direction. Here we limit the discussion to cases where atomic density varies with the constraint that OD is fixed. Once an experimental setup is complete, OD is a fixed number and the atomic density can be inferred through OD.  We are also particularly interested in low parametric gain regime where the pump laser is weak and far-detuned as illustrated in Fig.~\ref{schematic}.

The atomic density is described as $N(z) = N_{0} f_N(z)$, where $N_{0}$ is the mean atomic density and $f_N(z)$ is the atomic density profile function ($\int_{-L/2}^{L/2} f_N(z)dz=L$). The optical depth is given as $\textrm{OD} = \int_{-L/2}^{L/2} N(z)\sigma_{13} dz = N_{0} \sigma_{13} L$, where $\sigma_{13}$ is the on-resonance absorption cross section at the anti-Stokes transition. The profile function $f_N(z)$ appears in both the linear and the third-order nonlinear susceptibilities (Eqs.~\eqref{chi_s_def}-\eqref{chi_as_3rd_def}). Write $\chi_{as}(z) = \bar{\chi}_{as} f_N(z)$, and $\chi^{(3)}(z) = \bar{\chi}^{(3)} f_N(z)$, with $\bar{\chi}_{as}$ and $\bar{\chi}^{(3)}$ being the parts that are independent of $z$, from Eq.~\eqref{F_deltaomega} we have
\begin{equation}\label{F_omega_nz0}
F(\varpi) =\bar{\chi}^{(3)} E_p E_c\int_{-\frac{L}{2}}^{\frac{L}{2}} dz f_N(z)e^{-i\int_0^z\Delta k(z^\p)dz^\p}.
\end{equation}
When the pump laser is weak and far-detuned, the Stokes field is weak and $\chi_s\approx 0$, the phase matching term $\Delta k(z^\p) \approx \overbar{\Delta k} f_N(z^\p)$,  with $\overbar{\Delta k}$ defined as $\overbar{\Delta k} \equiv \w_{as0}/(2c)\ \bar{\chi}_{as}$, which does not vary with $z$.  Equation \eqref{F_omega_nz0} becomes
\begin{equation}\label{F_omega_nz}
F(\varpi) =\bar{\chi}^{(3)} E_p E_c\int_{-\frac{L}{2}}^{\frac{L}{2}} dz G^\p(z) e^{-i\overbar{\Delta k} G(z)},
\end{equation}
where
\begin{equation}
G(z) \equiv \int_0^z f_N(z^\p)dz^\p,
\end{equation}
and $G^\p(z)$ is its derivative.
After performing the integration on $z$ in Eq.~\eqref{F_omega_nz} and taking into account that the OD is fixed (or $\int_{-L/2}^{L/2}f_N(z)dz = L)$, and ignoring the vacuum phase mismatching term [$(\Delta k)_{vacuum} \approx 0$], we have
\begin{equation} \label{F_Q}
\begin{split}
F(\varpi) = &\bar{\chi}^{(3)} E_p E_c L\ \textrm{sinc}\left(\frac{\overbar{\Delta k} L}{2}\right) e^{i\overbar{\Delta k}(L/2-\beta)}, \\
Q(\varpi) = & e^{i(k_{s0} + k_{as0})L/2} e^{i\overbar{\Delta k}\beta},
\end{split}
\end{equation}
where $\beta \equiv \int_0^{L/2} f_N(z^\p)dz^\p$. Comparing Eqs.~\eqref{biphoton_1}, \eqref{psi_no_z} and \eqref{F_Q}, we find that $\psi(\tau)$ is the same as that when the atomic density is uniform.

Let's try to understand intuitively why the biphoton waveform is independent of the atomic distribution. The biphoton waveform depends on two factors: (1) the probability of biphoton generation by atoms in the atomic cloud, and (2) the time required for the photons to propagate through the atomic cloud to the detector. When the atomic distribution changes from uniform to non-uniform with a distribution, the probability of emitting photon pairs should follow this distribution. That is, in space where there are more atoms, the probability of emitting photons pairs increases. At the same time, the group velocity in this densely-populated space decreases and thus photons need more time to travel through this space to reach the detector. As the coincidence counting rate is in fact the probability of generating photon pairs divided by the time for the photons to reach the detector, the effect of longer travel time washes out that of the larger probability density. As a result, the biphoton waveform is not sensitive to the atomic density distribution profile.

Note that the above discussion is only valid when (1) atomic density varies in the longitudinal direction with the constraint that OD is fixed and (2) the pump laser is weak and far-detuned. Both conditions are of interests to our on-going experiments. If (1) is not satisfied but (2) is, our numerical analysis shows that the waveform will still be rectangle-like as in the case of uniform atomic density. But the group delay time is changed as it is determined by
$\tau_g = (2\gamma_{13}/|\Omega_c^2)\mathrm{OD}$~\cite{Du}.

%-------------------------------------------------------------------------------
\subsubsection{Pump laser has a $z$ profile} \label{section:pump_z}

If the pump laser has a non-uniform $z$ profile, namely, $E_p(z) = E_p f_p(z)$, the coupling laser has a uniform profile and the atomic cloud is homogeneous, Eq.~\eqref{F_deltaomega} becomes
\begin{equation} \label{F_deltaomega_1}
F(\varpi) = E_pE_c\chi^{(3)}(\varpi)\int_{-\frac{L}{2}}^\frac{L}{2} dz
                             f_p(z) e^{-i\int_0^z \Delta k(z^\p)dz^\p}.
\end{equation}
Here $\chi^{(3)}(\varpi)$ does not depend on the pump profile [see Eq.~\eqref{chi_as_3rd_def}].
When the pump laser is weak and far-detuned, the linear susceptibility of the Stokes field $\chi_s \approx 0$, and
\begin{equation}\label{delta_k_vg}
\Delta k(z^\p) \approx \frac{\varpi }{V_g},
\end{equation}
where  $V_g $ is the group velocity of the anti-Stokes photons. Now Eq.~\eqref{Q_omega} can be approximated as
\begin{equation}\label{Q_omega_2}
Q(\varpi) \approx e^{i\left(k_{s0} + k_{as0}\right)\frac{L}{2}} e^{i\varpi L/(2V_g)},
\end{equation}
where $k_{s0} = \w_{s0}/c$ is the central wavenumber of the Stokes photons and $k_{as0} = \w_{as0}/c$ is the central wavenumber of the anti-Stokes photons. Eq.~\eqref{F_deltaomega_1} at the same time can be approximated by
\begin{equation} \label{F_omega_2}
F(\varpi) \approx \chi^{(3)}E_pE_c\int_{-\frac{L}{2}}^\frac{L}{2} dz
																		f_p(z) e^{-i \varpi z/V_g }.
\end{equation}
Substituting Eqs.~\eqref{Q_omega_2} and \eqref{F_omega_2} into Eq.~\eqref{biphoton_1} results in
\begin{equation}\label{psi_coupling_z}
\begin{split}
\psi(\tau) \approx \frac{\sqrt{ \w_{s0}\w_{as0} } } {i4\pi c} E_p E_c e^{i\left(k_{s0} + k_{as0}\right)\frac{L}{2}}\\
\times\int_{-\frac{L}{2}}^\frac{L}{2} dz f_p(z)
                      \widetilde{\chi^{(3)}}\left(\tau+\frac{z}{V_g}-\frac{L}{2V_g}\right).              									
\end{split}
\end{equation}
Here
\begin{equation}\label{eq1}
\widetilde{\chi^{(3)}}\left(\tau+\frac{z}{V_g}-\frac{L}{2V_g}\right) \equiv \int d\varpi \ \chi^{(3)}(\varpi)
                                                  e^{-i\varpi\left(\tau+\frac{z}{V_g}-\frac{L}{2V_g}\right)}.
\end{equation}
With the change of integration variable from $z$ to $t$ as $t = \tau + z/V_g -L/(2V_g)$, integration in space domain in Eq.~\eqref{psi_coupling_z} changes to integration in time domain:
\begin{equation}\label{psi_coupling_z_1}
\begin{split}
& \psi(\tau) = \frac{\sqrt{ \w_{s0}\w_{as0} } } {i4\pi c} E_p E_c V_g e^{i\left(k_{s0} + k_{as0}\right)\frac{L}{2}} \\
& \times\int_{\tau-\frac{L}{V_g}}^\tau dt
                          f_p\left(\frac{L}{2} + V_g(t-\tau)\right) \widetilde{\chi^{(3)} } (t).
\end{split}									
\end{equation}
This is a convolution of the pump laser profile in space domain $f_p(z)$ and the third-order susceptibility in time-domain. In the group delay regime where the EIT window is much narrower than the $\chi^{(3)}$ spectrum, we can approximate $\chi^{(3)}(\varpi)\simeq\chi^{(3)}(0)$ in the integral \eqref{eq1}. Then Eq. \eqref{psi_coupling_z_1} reduces to
\begin{equation}\label{psi_coupling_z_2}
\begin{split}
& \psi(\tau) = \frac{\sqrt{ \w_{s0}\w_{as0} } } {i2 c} E_p E_c \chi^{(3)}(0) V_g e^{i\left(k_{s0} + k_{as0}\right)\frac{L}{2}} f_p\left(\frac{L}{2} - V_g\tau\right).
\end{split}									
\end{equation}
The argument in function $f_p(\frac{L}{2} - V_g\tau)$ suggests that the space-domain function $f_p(z)$ is mapped to the time-domain function $\psi(\tau)$ scaled with the anti-Stokes group delay.

Note that if the space-domain function $f_p(z)$ is a rectangular function, that is, when the pump power is uniform, we will have a time-domain rectangular $\psi(\tau)$.

The assumption that $\Delta k$ can be approximated by Eq.~\eqref{delta_k_vg} is valid when the loss in the medium is negligible. To account for the loss, we have to include an imaginary part $\alpha$ to anti-Stokes wavenumber as $k_{as} \approx k_{as0} + \varpi/V_g + i\alpha$. This imaginary part will then appear in $Q(\varpi)$ as an exponential decay factor $\exp{(-\alpha L/2)}$.

%-----------------------------------------------------
\subsubsection{Coupling laser has a $z$ profile}

If the coupling laser has a non-uniform profile in $z$ direction, the Rabi frequency, $\Omega_c(z) \propto E_c(z)$, depends on $z$. As a result, the linear and third-order responses of the atomic cloud depend on $z$ (see Eqs.~\eqref{chi_s_def}-\eqref{chi_as_3rd_def}), as well as the wavevectors. The analytical solution is too complicated. We will discuss the numerical results in Sec.~\ref{Numerical_result}.

%---------------------------------------------------------------------------------------------------------------------------
\subsection{Heisenberg Picture}

Now we turn to the Heisenberg picture where the vacuum state vector is time-invariant and the system is described by the evolution of the Stokes and anti-Stokes field operators. The coupling of the Stokes and anti-Stokes fields to the environment is included through Langevin force operators. In this picture, the space- and time-dependent Stokes and anti-Stokes field operators can be expressed as
\begin{equation}\label{efield}
\begin{split}
E_s^{(+)}(z,t) = & \sqrt{\frac{2\hbar \w_{s0}}{c\varepsilon_0 A}} \hat{a}_s(z,t) e^{i(-k_{s0} z - \w_{s0} t)}  \\
E_{as}^{(+)}(z,t) = & \sqrt{\frac{2\hbar \w_{as0}}{c\varepsilon_0 A}}\hat{a}_{as}(z,t) e^{i(k_{as0} z - \w_{as0} t)}
\end{split}
\end{equation}
The slowly-varying envelope field operators $\hat{a}_s(z,t)$ and $\hat{a}_{as}(z,t)$ in time domain are related to the frequency-domain operators $\hat{a}_s(z,-\varpi)$ and $\hat{a}_{as}(z,\varpi)$ through Fourier transform
\begin{equation}
\begin{split}
\hat{a}_s(z,t) = & \frac{1}{\sqrt{2\pi}}\int d\varpi\ \hat{a}_s(z,\varpi) e^{-i\varpi t}  \\
\hat{a}_{as}(z,t) = & \frac{1}{\sqrt{2\pi}}\int d\varpi\ \hat{a}_{as}(z,\varpi) e^{-i\varpi t},
\end{split}
\end{equation}
which is governed by the coupled Heisenberg-Langevin equations under the slowly varying envelope approximation
\begin{equation}\label{ode_omega}
\begin{split}
& \frac{\partial \hat{a}_{as}(z,\varpi)}{\partial z} + (\alpha_{as} - i \frac{\Delta k_0}{2}) \hat{a}_{as}(z,\varpi)
 - \kappa_{as} \hat{a}_s^\dag \left(z,-\varpi\right) = \hat{F}_{as} \\
& \frac{\partial  \hat{a}_s^\dag (z,-\varpi)}{\partial z} + (g_s + i \frac{\Delta k_0}{2}) \hat{a}_s^\dag (z,-\varpi)
  - \kappa_s \hat{a}_{as}(z,\varpi) = \hat{F}_s^\dag.
\end{split}
\end{equation}
Here
\begin{equation}\label{alpha_g}
\begin{split}
& \alpha_{as}(z,\varpi) =    -  \frac{i\w_{as}}{2c} \chi_{as}(z,\omega_{as0}+\varpi), \\
& g_{s}(z,\varpi) =          - \frac{i\w_s}{2c} \chi_s^{*}(z,\omega_{s0}-\varpi), \\
& \kappa_{as}(z,\varpi) =    \frac{i \w_{as}}{2c} \chi_{as}^{(3)}(z,\omega_{as0}+\varpi) E_p(z) E_c(z), \\
& \kappa_{s}(z,\varpi) =     \frac{i \w_s}{2c} \chi_s^{(3)*}(z,\omega_{s0}-\varpi) E_p^*(z) E_c^*(z),
\end{split}
\end{equation}
$F_{as}$ and $F_s^\dag$ are the contributions from the Langevin noises. They are given by
\begin{equation}
\begin{split}
\hat{F}_{as} & = \beta_{21}^{as} \hat{f}_{\sigma_{21}^\dag} + \beta_{24}^{as} \hat{f}_{\sigma_{24}^\dag}
         + \beta_{31}^{as} \hat{f}_{\sigma_{31}^\dag} + \beta_{34}^{as} \hat{f}_{\sigma_{34}^\dag},\\
 \hat{F}_s^\dag & = \beta_{21}^{s} \hat{f}_{\sigma_{21}^\dag} + \beta_{24}^{s} \hat{f}_{\sigma_{24}^\dag}
         + \beta_{31}^{s} \hat{f}_{\sigma_{31}^\dag} + \beta_{34}^{s} \hat{f}_{\sigma_{34}^\dag}.
\end{split}
\end{equation}
Here $\hat{f}_{\alpha^\dag_i}$ are Langevin force operators, $\beta_{\alpha_i}^{as}$ and $\beta_{\alpha_i}^{s}$ ($\alpha_i = 21,24,31,34$) are the noise coefficients. Detailed expressions for the noise coefficients are given in the Appendix.
$\Delta k_0 \equiv  k_{as0} - k_{s0}- (k_c - k_p)\cos\theta$. The expressions for susceptibilities
$\chi_{as}(z,\omega_{as0}+\varpi)$, $\chi_{s}(z,\omega_{s0}-\varpi)$, $\chi_{as}^{(3)}(z,\omega_{as0}+\varpi)$, and $\chi_{s}^{(3)}(z,\omega_{s0}-\varpi)$ are given in Sec.~\ref{Numerical_result} [Eqs.~\eqref{chi_s_def} to \eqref{chi_as_3rd_def}]. Defining $l \equiv L/2$, we have the following boundary conditions (vacuum at $z=\pm l$) for the coupled differential equations~\eqref{ode_omega}:
\begin{equation}\label{Boundarycondition1}
[\hat{a}_s(l,\varpi), \hat{a}^\dag_s(l,\varpi^\p)] = [\hat{a}_{as}(-l,\varpi), \hat{a}^\dag_{as}(-l,\varpi^\p)] = \delta(\varpi - \varpi^\p),
\end{equation}
\begin{equation}\label{Boundarycondition2}
\langle\hat{a}^\dagger_s(l,\varpi)\hat{a}_s(l,\varpi)\rangle = \langle\hat{a}^\dagger_{as}(-l,\varpi) \hat{a}_{as}(-l,\varpi)\rangle = 0.
\end{equation}

As shown in the Appendix, the general solution to Eqs.~\eqref{ode_omega} at the output surface can be written as
\begin{equation}\label{0L}
\begin{split}
&\left( \begin{array}{c} \hat{a}_{as}(l,\varpi) \\ \hat{a}_s^\dag (-l,-\varpi)  \end{array} \right)
= \begin{pmatrix}
     A(\varpi) & B(\varpi) \\
		 C(\varpi) & D(\varpi)
	\end{pmatrix}
  	\left( \begin{array}{c} \hat{a}_{as}\left(-l,\varpi\right) \\ \hat{a}_s^\dag \left(l,-\varpi\right)  \end{array} \right) \\
	& + \sum_{\alpha_i}\int_{-l}^l dz \begin{pmatrix} P_{\alpha_i} \\ Q_{\alpha_i} \end{pmatrix} \hat{f}_{\alpha_i^\dag}.
\end{split}
\end{equation}
$P_{\alpha_i}$ and $Q_{\alpha_i}$ ( $\alpha_i= 21, 24, 31, 34$ ) are functions of $z$ and $\varpi$, and contain contributions from Langevin noises (see Appendix for their expressions).

The Glauber correlation function can be calculated from
\begin{eqnarray}
 G^{(2)}_{s,as}(\tau) &\equiv & \langle\hat{a}^{\dagger}_{as}(l,t_{s}+\tau)\hat{a}^{\dagger}_{s}(-l,t_{s})\hat{a}_{s}(-l,t_{s})\hat{a}_{as}(l,t_{s}+\tau)\rangle \nonumber\\
 &=&  \left| \psi(\tau) \right|^2 + R_{as}R_s.\label{GlauberCorrelationH}
\end{eqnarray}
Here the Stokes and anti-Stokes two-photon relative wave amplitude on the output surface ($z=\pm l$) is
\begin{equation}\label{psi_heisenberg}
\begin{split}
 &\psi(\tau \equiv t_{as}-t_{s})\equiv\langle\hat{a}_{s}(-l,t_s)\hat{a}_{as}(l,t_{as})\rangle \\
  &=  \int{ \frac{d\varpi}{2\pi} e^{-i\varpi\tau}  \Big[ B\left(\varpi\right) D^*\left(\varpi\right)
		                                + \sum_{\alpha_i,\alpha_j}\int_{-l}^l dz
																		      Q^*_{\alpha_i}D_{\alpha_i,\alpha_j^\dag}P_{\alpha_j}\Big]
									                }.
\end{split}
\end{equation}
The generation rates of the Stokes photons and anti-Stokes photons are given by
\begin{equation}\label{PhotonRates}
\begin{split}
 & R_s\equiv\langle\hat{a}^\dagger_{s}(-l,t_s)\hat{a}_{s}(-l,t_s)\rangle \\
& =\int{ \frac{d\varpi}{2\pi}\Big( |C(\varpi)|^2 + \sum_{\alpha_i,\alpha_j}\int_{-l}^l\ dz\,
																		      Q_{\alpha_i}D_{\alpha_i^\dag,\alpha_j}Q^*_{\alpha_j}\Big)
											}, \\
& R_{as}\equiv\langle\hat{a}^\dagger_{as}(l,t_{as})\hat{a}_{as}(l,t_{as})\rangle \\
&= \int \frac{d\varpi}{2\pi} \Big(|B(\varpi)|^2 + \sum_{\alpha_i,\alpha_j}\int_{-l}^l dz\,
																		      P^*_{\alpha_i}D_{\alpha_i,\alpha_j^\dag}P_{\alpha_j}\Big),
\end{split}
\end{equation}
respectively. $R_{as}R_s$ term in Eq.~\eqref{GlauberCorrelationH} results from accidental coincidence between uncorrelated photons because the photon pairs are produced stochastically and the time separation between different pairs is unpredictable. Detailed derivation is given in the Appendix.

The photon pair generation rate can be computed as
\begin{equation}
\begin{split}
& R = \int |\psi(\tau)|^2 d\tau \\
& = \int \frac{d\varpi}{2\pi}\Big|B\left(\varpi\right) D^*\left(\varpi\right)
                                            + \sum_{\alpha_i,\alpha_j}\int_{-l}^l dz\,
																		      Q^*_{\alpha_i}D_{\alpha_i,\alpha_j^\dag}P_{\alpha_j}\Big|^2.
\end{split}
\end{equation}
This is the area under the Stokes--anti-Stokes correlation function minus the uncorrelated background.

%%%%%%%%%%%%%%%%%%%%%%%%%%%%%--- AC ---%%%%%%%%%%%%%%%%%%%%%%%%%%%%%%%%%%%%%%%%%%%%%%%%%%%%%%%%%%%%%%%%%%%%%%%
Note that we can also define anti-Stokes and Stokes biphoton amplitude as
\begin{equation}\label{psi_p_heisenberg}
\begin{split}
 &\psi(\tau)\equiv\langle\hat{a}_{as}(l,t_{as})\hat{a}_{s}(-l,t_s)\rangle \\
  &=  \int{ \frac{d\varpi}{2\pi} e^{-i\varpi\tau}  \Big[ A\left(\varpi\right) C^*\left(\varpi\right)
		                                + \sum_{\alpha_i,\alpha_j}\int_{-l}^l dz
																		      P_{\alpha_i}D_{\alpha_i^\dag,\alpha_j}Q^*_{\alpha_j}\Big]
									                }.
\end{split}
\end{equation}
%%% check the formula

With the contribution from Langevin noise, Eqs~\eqref{psi_heisenberg} and \eqref{psi_p_heisenberg} should give the same results numerically. This have been verified by our numerical calculations with a wide range of parameters. Note that  when the pump is weak and far-detuned, the majority of the atomic population is the ground state. The diffusion coefficients $D_{\alpha_i,\alpha_j^\dag}$, which appears in Eq.~\eqref{psi_heisenberg}, are very small as they only depend on the excited states population (see Appendix for details). This makes the contribution from Langevin noise to Eq~\eqref{psi_heisenberg} negligible. However, the diffusion coefficients $D_{\alpha_i^\dag,\alpha_j}$, which appears in Eq.~\eqref{psi_p_heisenberg}, are large as they also depend on the ground state population. As a result, the contribution from Langevin noise to Eq.~\eqref{psi_p_heisenberg} is large. Therefore in the following discussion, for convenience, we take the following approximation to analyze the bphoton temporal wave function
\begin{equation}\label{psiapp_heisenberg}
\psi(\tau)\simeq \int{ \frac{d\varpi}{2\pi} B\left(\varpi\right) D^*\left(\varpi\right) e^{-i\varpi\tau}}.
\end{equation}

%%%%%%%%%%%%%%%%%%%%%%%%%%%%%%%%%%%%%%%%%%%%%%%%%%%%%%%%%%%%%%%%%%%%%%%%%%%%%%%%%%%%%%%%%%%%%%%%%%%%%%%%%%%%
The normalized cross-correlation function of Stokes--anti-Stokes photons is
\begin{equation}
\begin{split}
g^{(2)}_{s,as}(\tau)
 & \equiv \frac{\langle\hat{a}^{\dag}_{as}(l,t_{s}+\tau)\hat{a}^{\dag}_{s}(-l,t_{s})\hat{a}_{s}(-l,t_{s})\hat{a}_{as}(l,t_{s}+\tau)\rangle} { \langle\hat{a}^{\dag}_{as}(l,t_{s}+\tau)\hat{a}_{as}(l,t_{s}+\tau)\rangle
           \langle\hat{a}^{\dag}_{s}(-l,t_{s})\hat{a}_{s}(-l,t_{s})\rangle } \\
& = \frac{G^{(2)}_{s,as}(\tau)}{R_{as}R_s} = 1+\frac{|\psi(\tau)|^2}{R_{as}R_s}.
\end{split}
\end{equation}
The normalized autocorrelation function of the anti-Stokes photons is
\begin{equation}\label{g2as}
\begin{split}
& g_{as,as}^{(2)}(\tau) = \frac{\langle \hat{a}_{as}^\dag(l,0) \hat{a}_{as}^\dag(l,\tau) \hat{a}_{as}(l,\tau) \hat{a}_{as}(l,0)\rangle}
   {\langle \hat{a}_{as}^\dag(l,0) \hat{a}_{as}(l,0)\rangle \langle \hat{a}_{as}^\dag(l,\tau) \hat{a}_{as}(l,\tau)\rangle}
	\\	
& = \frac{\bigg|\displaystyle\int\frac{d\varpi}{2\pi} e^{-i\varpi \tau}
                                  \Big(|B(\varpi)|^2 + \sum\limits_{\alpha_i,\alpha_j}
                                      \int_{-l}^l dz
																		      P^*_{\alpha_i}D_{\alpha_i,\alpha_j^\dag}P_{\alpha_j}\Big)\bigg|^2}
				{\bigg[ \displaystyle\int \frac{d\varpi}{2\pi} \Big(|B(\varpi)|^2
												                + \sum\limits_{\alpha_i,\alpha_j}
																				\int_{-l}^l dz
																		      P^*_{\alpha_i}D_{\alpha_i,\alpha_j^\dag}P_{\alpha_j}\Big)\bigg]^2}
  \\
	& \ + 1,
\end{split}
\end{equation}
and of the Stokes photons is,
\begin{equation}\label{g2s}
\begin{split}
& g_{s,s}^{(2)}(\tau) = \frac{\langle \hat{a}_{s}^\dag(l,0) \hat{a}_{s}^\dag(l,\tau) \hat{a}_{s}(l,\tau) \hat{a}_{s}(l,0)	\rangle}
    {\langle \hat{a}_{s}^\dag(l,0) \hat{a}_{s}(l,0)\rangle \langle \hat{a}_{s}^\dag(l,\tau) \hat{a}_{s}(l,\tau)\rangle}
\\
& = \frac{\bigg|\displaystyle\int \frac{d\varpi}{2\pi} e^{-i\varpi \tau}\Big(|C(\varpi)|^2 + \sum\limits_{\alpha_i,\alpha_j}
                                      \int_{-l}^l dz
																		      Q_{\alpha_i}D_{\alpha_i^\dag,\alpha_j}Q^*_{\alpha_j}\Big)
																					\bigg|^2}
						{\bigg[ \displaystyle\int \frac{d\varpi}{2\pi} \Big(|C(\varpi)|^2 + \sum\limits_{\alpha_i,\alpha_j}
						                          \int_{-l}^l dz
																		      Q_{\alpha_i}D_{\alpha_i^\dag,\alpha_j}Q^*_{\alpha_j}\Big) \bigg]^2}
\\
& \ + 1.
\end{split}
\end{equation}
It is clear that Eqs. \eqref{g2as} and \eqref{g2s} are the autocorrelation functions for multimode chaotic light sources with $g_{as,as}^{(2)}(0)=g_{s,s}^{(2)}(0)=2$. For classical light, there is the Cauchy-Schwarz inequality $\left[g^{(2)}_{s,as}(\tau)\right]^2/[g^{(2)}_{s,s}(0)g^{(2)}_{as,as}(0)] \le 1$~\cite{Clauser1974}. Therefore violation of the Cauchy-Schwarz inequality is a measure of the nonclassical property of the biphoton source, which requires
$[g^{(2)}_{s,as}(\tau)]_{max}>2.$

\subsubsection*{No $z$ dependence} \label{Heisenberg_noZ}

When the atomic density is homogenous, the pump and coupling laser beams have uniform intensities in the atomic cloud, i.e., when $N(z) = N_0$, $E_p(z) = E_p$, $E_c(z) = E_c$, there will be no $z$ dependence for $\alpha_{as}$, $g_s$, $\kappa_{as}$ and $\kappa_s$ as well in Eq.~\eqref{ode_omega}. In this case the coupled equation \eqref{ode_omega} can be solved analytically. The result is
\begin{equation}\label{BL_const}
B(\varpi) = \frac{2\kappa_{as}}
            { q+Q \coth \left(lQ \right)
						}, 							
\end{equation}
\begin{equation}\label{D0_const}
D(\varpi) = \frac{Q \exp[(g_s +\alpha_{as})l]}
            { q\sinh\left(lQ\right)
						   + Q\cosh \left(lQ\right)
						}.
\end{equation}
Here $q \equiv \alpha_{as} - g_s - i\Delta k_0$, which depends on the linear response of the medium,  and $ Q \equiv \sqrt{q^2+4 \kappa_s \kappa_{as}}$. Note that $\alpha_{as}$, $g_s$, $\kappa_{as}$ and $\kappa_s$ are still functions of $\varpi$.
Here we did not include the contribution from the Langevin operators. The reason is: when we limit our discussion to a weak and far-detuned pumping and therefore majority of the atomic population is in the ground state. In this case the contribution from Langevin noise operators to $B(\varpi)$ and $D(\varpi)$ is very small. This is confirmed by our numerical analysis. Please see Appendix for detailed discussion.

In the limit of low parametric gain where $4\kappa_s\kappa_{as}\ll q^2$, Eqs.~\eqref{BL_const} and \eqref{D0_const} reduce to
\begin{equation}\label{BL_const_1}
B\left(\varpi\right) = \frac{2\kappa_{as}} { q\left[1+\coth(ql)\right]}, 							
\end{equation}
and
\begin{equation}\label{D0_const_1}
D\left(\varpi\right) = e^{(2g_s + i\Delta k_0)l},
\end{equation}
respectively.

To compare with the result in the interaction picture, we need to write $q$ and $g_s$ in terms of the Stokes and anti-Stokes wave numbers in the medium ($k_s$ and $k_{as}$).  The anti-Stokes wavenumber in the medium is $k_{as} \approx \w_{as0}/c\ (1+\chi_{as}/2) = k_{as0} + \Delta k_{as}$, with $\Delta k_{as} \equiv \w_{as0} \chi_{as}/(2c)$,
and the Stokes wavenumber in the medium is $k_{s} \approx \w_{s}/c\ (1+\chi_{s}/2) = k_{s0} + \Delta k_{s}$, with $\Delta k_{s} \equiv \w_{s0} \chi_{s}/(2c)$. Then $q\approx -i(\Delta k_{as} - \Delta k_s^* + \Delta k_0)$ and
\begin{eqnarray}
B\left(\varpi\right)D^*(\varpi) &= & L\kappa_{as} \textrm{sinc}\left[(\Delta k_{as} - \Delta k_s^{*} + \Delta k_0)l\right] \nonumber\\
&\times&   e^{i(\Delta k_{as} - \Delta k_s^{*} + 2\Delta k_s)l}.
\end{eqnarray}
If the imaginary part of $\Delta k_s$ is small, or Raman gain is small, $\Delta k_s^{*} \approx \Delta k_s$. The product $B(\varpi)D^*(\varpi)$ becomes
\begin{eqnarray}
B\left(\varpi\right)D^*\left(\varpi\right) &= & L\kappa_{as} \textrm{sinc}\left((\Delta k_{as} - \Delta k_s + \Delta k_0)l\right) \nonumber \\
&\times& e^{i(\Delta k_{as} + \Delta k_s)l}.
\end{eqnarray}
The argument inside the 'sinc' function can be rewritten as $\Delta k_{as} - \Delta k_s + \Delta k_0 \equiv \Delta k$. The biphoton wavefunction is now
\begin{equation}\label{H_psi_const}
\begin{split}
& \psi(\tau) =  - \frac{\omega_{as0}}{i 4\pi c} E_p E_c L e^{-i(k_{as0}+k_{s0})L/2}  \\
&  \times  \int d\varpi \chi^{(3)}_{as}(\varpi)\textrm{sinc}\left(\Delta k L/2\right)e^{i(k_{as} + k_s)L/2} e^{-i\varpi \tau}.
\end{split}
\end{equation}
Comparing Eqs.~\eqref{biphoton_1}, \eqref{F_deltaomega}, and \eqref{Q_omega} with \eqref{H_psi_const}, and taking into account that $\w_{s0} \approx \w_{as0}$, we obtain the same $|\psi(\tau)|$ as that in the interaction picture.

%---------------------------------------------------------------------------------------------------------------------------
%\subsubsection{Atom density has a $z$ profile}

%In Maxwell Heisenberg-Langevin approach, atom density $N(z) = N_0 f(z)$ enters the coupled Eq.~\eqref{ode_omega} through %$\alpha_{as}$, $g_s$, $\kapps_{as}$ and $\kappa_s$. For a $\Delta_k \sim 0$, the density profile $f(z)$ eventually% %changes the

%---------------------------------------------------------------------------------------------------------------------------
\section{Numerical Results} \label{Numerical_result}

We take $^{85}$Rb cold atomic ensemble for numerical simulations. The relevant atomic energy levels involved are $|1\rangle=|5S_{1/2},F=2\rangle$,
$|2\rangle=|5S_{1/2},F=3\rangle$, $|3\rangle=|5P_{1/2},F=3\rangle$, and $|4\rangle=|5P_{3/2},F=3\rangle$. The pump laser is detuned by $\Delta_p=2\pi\times 150$~MHz. The atomic medium has a length L=2 cm.

We work in the ground-state approximation where the field of the pump laser is weak and far-detuned from $|1\rangle \to |4\rangle$ transition so that most of the atomic population is in the ground state. The linear and third-order susceptibilities of the Stokes and anti-Stokes fields are~\cite{Wen2006, Braje, WenRubin}
\begin{widetext}
\begin{align}
& \chi_s(z,\omega_{s0}-\varpi) = \frac{N(z)|\mu_{24}|^2 |\Omega_p(z)|^2 (\varpi-i\gamma_{13}) / (\varepsilon_0\hbar)}
                               { (\Delta_p^2+\gamma_{14}^2)
													                [|\Omega_c(z)|^2-4(\varpi-i\gamma_{12})(\varpi-i\gamma_{13})]} \ ,
																					\label{chi_s_def}\\
& \chi_{as}(z,\omega_{as0}+\varpi) = \frac{4 N(z)|\mu_{13}|^2  (\varpi+i\gamma_{12}) / (\varepsilon_0\hbar)}
                                  { |\Omega_c(z)|^2-4(\varpi+i\gamma_{12})(\varpi+i\gamma_{13})} \ ,
	\label{chi_as_def}\\
& \chi_s^{(3)}(z,\omega_{s0}-\varpi) = \frac{N(z)\mu_{13}\mu_{24}\mu_{14}^*\mu_{23}^* / (\varepsilon_0\hbar^3)}
                               { (\Delta_p +i\gamma_{14})
													                 \left[|\Omega_c(z)|^2-4(\varpi-i\gamma_{12})(\varpi-i\gamma_{13})\right]} \ ,  \label{chi_s_3rd_def}\\
& \chi_{as}^{(3)}(z,\omega_{as0}+\varpi) = \frac{N(z)\mu_{13}\mu_{24}\mu_{14}^*\mu_{23}^* / (\varepsilon_0\hbar^3) }
                               {(\Delta_p +i\gamma_{14})
													                 \left[|\Omega_c(z)|^2-4(\varpi+i\gamma_{12})(\varpi+i\gamma_{13})\right]} \ .
	\label{chi_as_3rd_def}
\end{align}
\end{widetext}
Here, $N(z)=N_0 f_N(z)$ is the MOT atomic density, $\mu_{ij}$ is the dipole moment for $|i\rangle$ to $|j\rangle$ transition,  $\Omega_p(z) =\Omega_p f_p(z)= \mu_{41}E_p(z)/\hbar$ is the Rabi frequency of the pump laser field, and $\Omega_c(z)=\Omega_c f_c(z)=\mu_{32}E_c(z)/\hbar$ is the Rabi frequency of the coupling laser field. $\gamma_{ij}$ is the dephasing rate between $|i\rangle$ and  $|j\rangle$. As the natural linewidth of $^{85}$Rb atoms is $\Gamma = 2\pi\times 6$~MHz, we have $\gamma_{13} = \gamma_{14} = \gamma_{23} = \Gamma/2$.  For simulation, we take the ground-state dephasing rate $\gamma_{12} = 0.01\gamma_{13}$. Other parameters are $OD=N_0 \sigma_{13} L=150$, $\Omega_p = 2\pi\times 1.2$~MHz and $\Omega_c = 2\pi\times 12$~MHz, unless they are specified.

Note that in calculating Stokes--anti-Stokes biphoton wavefunction in the interaction picture, we take $\chi^{(3)}(\varpi) = \chi_{as}^{(3)}(\omega_{as0}+\varpi)=\chi_{s}^{(3)}(\omega_{s0}+\varpi)$.

\subsection{Photon Properties}

%--------------------------------------------------
\begin{figure}[h]
\centering
\includegraphics[width=\linewidth]{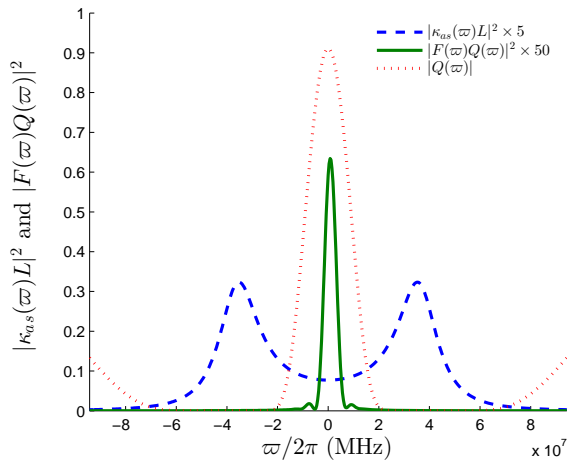}
\caption{(Color online) $|\kappa_{as}(\varpi)L|^2$, $Q(\varpi)$ and two-photon spectrum $|F(\varpi)Q(\varpi)|^2$ vs the varying component of the frequency of the anti-Stoke field $\varpi$. Note that $|\kappa_{as}(\varpi)L|^2$ is magnified 5 times and $|F(\varpi)Q(\varpi)|^2$ is magnified 50 times in the figure.}
\label{spectrum}
\end{figure}

In Sec.~\ref{Heisenberg_noZ}, we proved that both the interaction and Heisenberg pictures give the same biphoton waveform, characterized by $\psi(\tau)$ when there is no $z$ dependence in the atomic density and the driving laser fields. The biphoton waveform is determined by two parts, $F(\varpi)$ and $Q(\varpi)$. $F(\varpi)$ involves the nonlinear response $\chi^{(3)}$ and the phase-mismatching effect, while $Q(\varpi)$ implies the linear propagation effect in the atomic medium. In the expression of $\chi^{(3)}$, the term $[|\Omega_c(z)|^2-4(\varpi+i\gamma_{12})(\varpi+i\gamma_{13})]$ in the denominator can be rewritten as $(-1/4)[(\varpi -\Omega_e/2+i\gamma_e)(\varpi +\Omega_e/2+i\gamma_e)]$, with $\Omega_e \equiv \sqrt{|\Omega_c|^2-(\gamma_{13}-\gamma_{12})^2}$ is the effective Rabi frequency, and $\gamma_e \equiv (\gamma_{12}+\gamma_{13})/2$ is the effective dephasing rate. It can be seen from the rewritten term that there are two resonances $\varpi = \pm \Omega_e/2$,  with linewidth determined by $2\gamma_e$. Inside $F(\varpi)$, there is also the term $\textrm{sinc}(\Delta k L/2)$, and its bandwidth is determined by the group-delay time, $\tau_g$, as $\Delta \w_g \equiv (2\pi\times 0.88)/\tau_g$, here $\tau_g = L/V_g \simeq (2 \gamma_{13} \textrm{OD} )/|\Omega_c|^2$~\cite{Du}. For $Q(\varpi)$, which determines the EIT transmission, its bandwidth can be calculated from Eqs.~\eqref{chi_s_def} and \eqref{chi_as_def}. It gives $\Delta \w_{tr} \simeq |\Omega_c|^2/(2 \gamma_{13} \sqrt{\textrm{OD}})$.

When $(\Omega_e, 2\gamma_e) < (\Delta\w_g, \Delta\w_{tr})$, the system is in the damped Rabi oscillation regime. This happens when $|\Omega_c|$ is large and the OD is small. The biphoton waveform is determined by the two resonances of the third-order response. When the OD is large (typically $\textrm{OD} \geq 4 \pi ^2$), we may have $\Omega_e > \Delta\w_{tr} > \Delta\w_g$. Then the two resonances are suppressed by the phase mismatching $\textrm{sinc}(\Delta k L/2)$ and the off-resonance EIT absorption. In this case, the phase-mismatching term sets the limit of the bandwidth of the biphotons and the group delay time is related to the biphoton correlation time. This is the group delay regime.

In Fig.~\ref{spectrum}, we plot $|\kappa_{as}(\varpi) \times L |^2$,  $|Q(\varpi)|$, and  $|F(\varpi) \times Q(\varpi)|^2$ as a function of $\varpi$. The function $|\kappa_{as}(\varpi) \times L |^2$ has two peaks, which are far-detuned from the central frequency of $|F(\varpi) \times Q(\varpi)|^2$ and $|Q(\varpi)|$. Note that $\kappa_{as}(\varpi)$ is given in Eq.~\eqref{alpha_g} and is proportional to $\chi^{(3)}$. Note also that as $\textrm{sinc}(\Delta k L/2)$ function has the same spectrum as that of $|F(\varpi) \times Q(\varpi)|^2$, it is not plotted in the figure. It is clear that Fig.~\ref{spectrum} lies in the group delay regime. In this paper, we limit our discussion to the group delay regime.

%--------------------------------------------------
\begin{figure}[h]
\centering
\includegraphics[width=\linewidth]{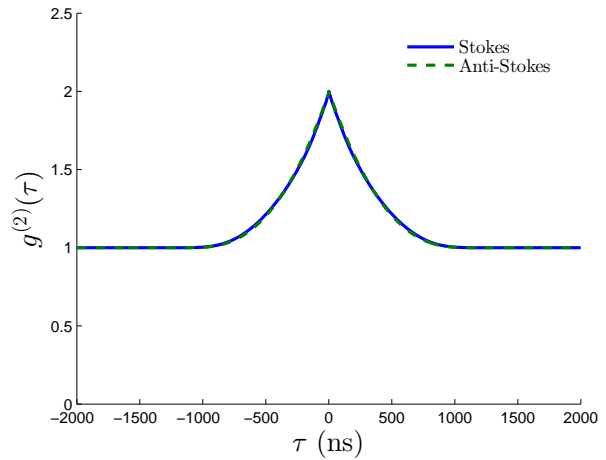}
\caption{(Color online) Normalized auto-correlation functions $g_{s,s}^{(2)}(\tau)$ and  $g_{as,as}^{(2)}(\tau)$ calculated in the Heisenberg picture.  No $z$ dependence in the atomic density and the driving laser fields. }
\label{g2_tau}
\end{figure}
%
%
%--------------------------------------------------
\begin{figure}[h]
\centering
\includegraphics[width=\linewidth]{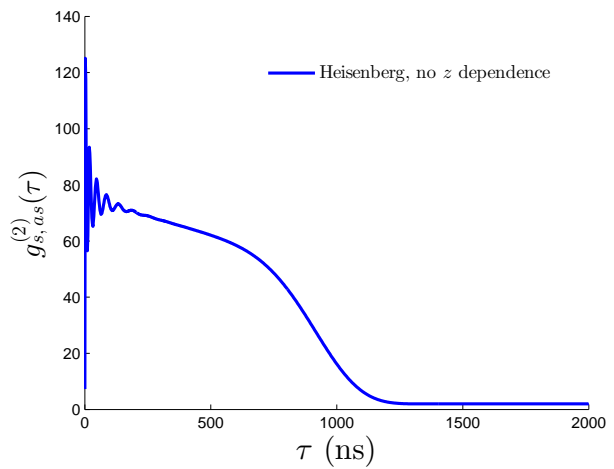}
\caption{(Color online) Normalized cross-correlation function $g_{s,as}^{(2)}(\tau)$ calculated in the Heisenberg picture. No $z$ dependence in atomic density and the driving laser fields.}
\label{g2_s_as}
\end{figure}
%------------------------------------------------------

When the parametric gain is small as in our case, the Stokes and anti-Stokes are generated spontaneously in pairs. The multimode chaotic nature is verified by their second-order coherence shown in Fig.~\ref{g2_tau}, as the normalized auto-correlation functions obtained in the Heisenberg picture: $g_{s,s}^{(2)}(0) = g_{as,as}^{(2)}(0) = 2$, and $1 \le g_{as,as}^{(2)}(\tau) \le 2$ as well as $1 \le g_{s,s}^{(2)}(\tau) \le 2$.
%
%---------------------------------------------------

The normalized cross-correlation $g_{s,as}^{(2)}(\tau)$ in Fig.~\ref{g2_s_as} shows a rectangular shape. The correlation time is nearly 1~$\mu$s, which is determined by the bandwidth of the biphoton spectrum $|F(\varpi) \times Q(\varpi)|^2$ in Fig.~\ref{spectrum}.

%--------------------------------------------------
\begin{figure}[h]
\centering
\includegraphics[width=\linewidth]{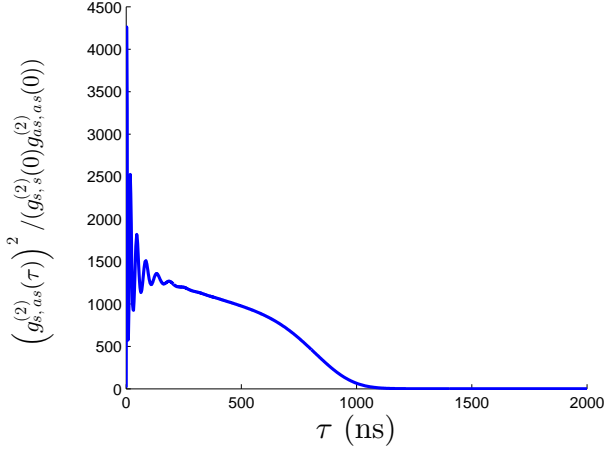}
\caption{(Color online) The ratio of normalized cross-correlation function over normalized auto-correlation function $\left(g_{s,as}^{(2)}(\tau)\right)^2/(g_{s,s}^{(2)}(0) g_{as,as}^{(2)}(0))$ calculated in the Heisenberg picture. No $z$ dependence in atomic density and the driving laser fields.}
\label{g2_ratio}
\end{figure}

%--------------------------------------------------
\begin{figure}[h]
\centering
\includegraphics[width=\linewidth]{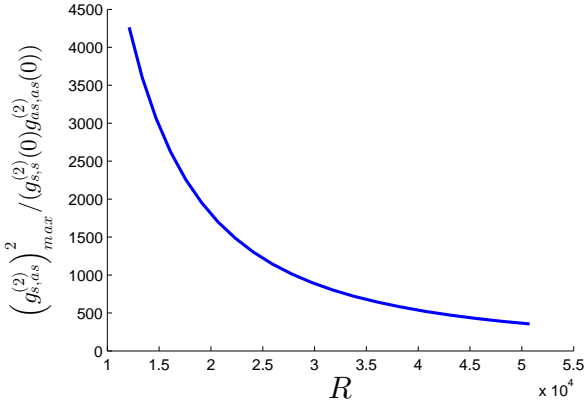}
\caption{(Color online) The ratio of normalized cross-correlation function over normalized auto-correlation function $\left(g_{s,as}^{(2)}\right)^2_{max}/(g_{s,s}^{(2)}(0) g_{as,as}^{(2)}(0))$ vs the photon pair emission rate $R$ calculated in the Heisenberg picture as the pump power doubles. No $z$ dependence in atomic density and the driving laser fields. The normalized cross-correlation function $g_{s,as}^{(2)}$ is taken at its maximum value. }
\label{g2_vs_R}
\end{figure}
%------------------------------------------------------

To determine the properties of the generated biphotons,  we calculate the ratio of the normalized cross-correlation function over the normalized auto-correlation function $[g_{s,as}^{(2)}(\tau)]^2/[g_{s,s}^{(2)}(0) g_{as,as}^{(2)}(0)]$.  As shown in Fig.~\ref{g2_ratio}, the Cauchy-Schwarz inequality is violated by a factor of about 4,200 and the biphoton nonclassical property is clearly confirmed.

Next, as we increase the pump power to increase the photon generation rate, the parametric gain increases, but the factor of violation of Cauchy-Schwarz inequality decreases, as shown in Fig.~\ref{g2_vs_R}.

Note that because the perturbation theory in the interaction picture describes only the two-photon process, the single photon generation rates $R_s$ and $R_{as}$ in the interaction picture cannot be described adequately by the biphoton state. As such, we obtain the normalized cross- and auto-correlation function $g_{s,as}^{(2)}(\tau)$, $g_{s,s}^{(2)}(\tau)$ and $g_{as,as}^{(2)}(\tau)$ in the Heisenberg picture.

%-------------------------------------------------------------------------------
\subsection{Comparison of the two formalisms}
%---------------------------------------------------------------------------------

We compare both models in the interaction and Heisenberg pictures by computing the second-order Glauber function
$G_{s,as}^{(2)}(\tau)$ numerically for cases with and without $z$ dependence in the atomic density, and the driving laser fields.

We have shown theoretically that when there is no $z$-dependence in the atomic density and the driving field intensities both models agree well when $q^2 \gg 4\kappa_s\kappa_{as}$, or when the linear response is much larger than the third-order response in the atomic medium. This is the low parametric gain regime. In the following figures, we show numerically that both models agree well in this regime when spatial dependence is absent or present in the the atomic density and driving laser fields.
%--------------------------------------------------
\begin{figure}[h]
%\vskip -1em
\centering
\includegraphics[width=\linewidth]{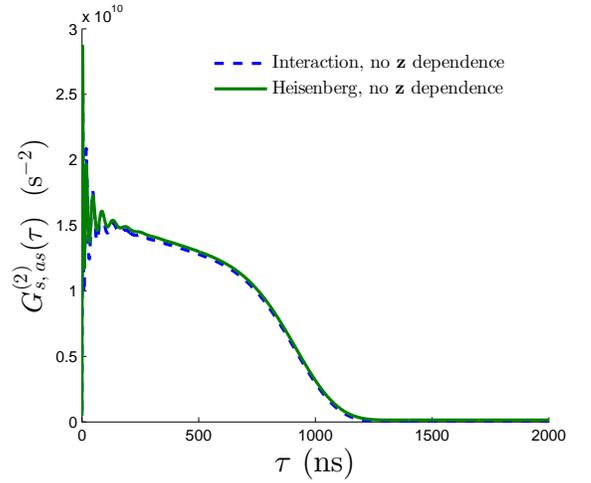}
%\vspace{-1em}
\caption{(Color online) $G_{s,as}^{(2)}(\tau)$ in both the interaction and Heisenberg pictures for uniform atomic medium, and uniform pump and coupling laser profiles in $z$ direction.}
\label{const_z}
\end{figure}
%---------------------------------------------------

Figure \ref{const_z} shows that with uniformly distributed atomic density and uniform pump and coupling field amplitudes in the $z$ direction, both models agree well in predicting the second-order Glauber function $G_{s,as}^{(2)}(\tau)$.  The curves are rectangle-like with an oscillatory optical precursor.

\begin{figure}[h]
\centering
\includegraphics[width=\linewidth]{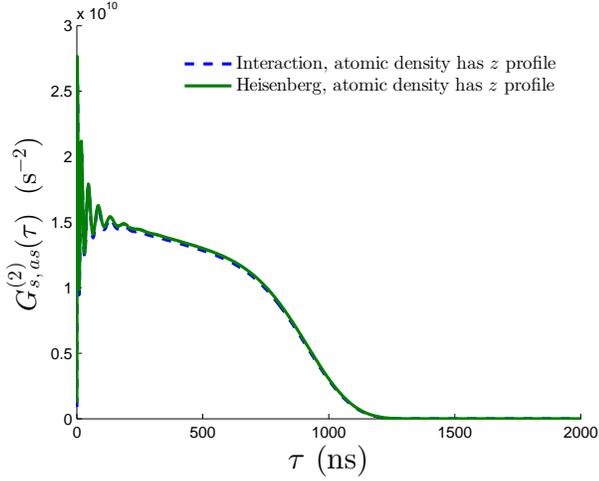}
\caption{(Color online) $G^{(2)}_{s,as}(\tau)$ in both the interaction and Heisenberg pictures for uniform pump and coupling laser profile in $z$ direction, but the atomic density is nonuniform with modulation function $f_N(z)=6(z+L/2)(L/2-z)/L^2 $.}
\label{N-varies-pc-const}
\end{figure}

Next we look at the case where the atomic density is not uniform in the $z$ direction, but is modulated in such a way that the total OD is unchanged, i.e., the modulation function $f_N(z)$ satisfies $\int_{-L/2}^{L/2} f_N(z) dz = L$. Figure \ref{N-varies-pc-const} shows that both models produce the same numerical results.
%---------------------------------------------------------------

\begin{figure}[h]
\begin{center}$
\begin{array}{c}
\includegraphics[width=\linewidth]{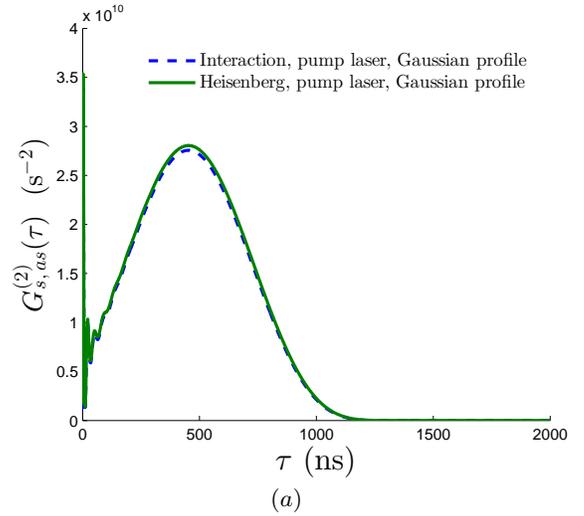}  \\
(a) \\
\includegraphics[width=\linewidth]{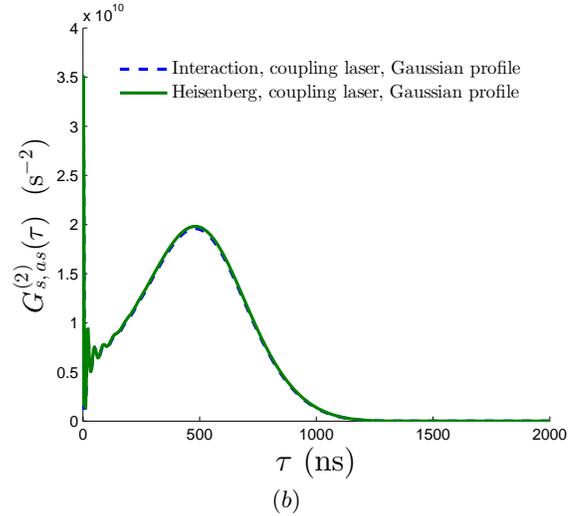} \\
(b)
\end{array}$
\end{center}
\caption{(Color online) $G^{(2)}_{s,as}(\tau)$ in both the interaction and Heisenberg pictures for uniform atom density and (a) only pump laser has a Gaussian profile in the $z$ direction, and (b) only coupling laser has Gaussian profile in the $z$ direction. The Gaussian modulation function is $f_{p,c}(z) = 1/0.65 \exp( -z^2/ (L/2)^2)$. }
\label{either-p-or-c}
\end{figure}

In Fig.~\ref{either-p-or-c}, we give a Gaussian profile to the pump or coupling laser such that $E_p(z) = E_p f_p(z)$ or
$E_c(z) = E_c f_c(z)$, with $f_p(z) = f_c(z) = 1/0.65 \exp( -z^2/ (L/2)^2)$.  Both models produce the same $G^{(2)}_{s,as}(\tau)$. Note that the shape is very different from those in Figs.~\ref{const_z} and \ref{N-varies-pc-const}. Here the waveform is not rectangle-like but Gaussian-like with a huge bump in the middle. This is a result of Eq.~\eqref{psi_coupling_z_1}, where the space-domain modulating function $f_p(z)$ determines the shape of the time-domain waveform.  This Gaussian-like biphoton waveform was observed in experiment by Zhao \textit{et al}~\cite{ZhaoOPTICA2014} for the first time, and was fitted well by a Gaussian modulation to the pump field. As discussed in Fig.~\ref{spectrum}, in the group delay regime, the third-order susceptibility is almost a constant in the biphoton frequency detuning window. When the atomic density (or OD) is high, the value of $\chi^{(3)}$ is large, the effect of modulation caused by the driving field profile is more significant. Therefore this Gaussian-like waveform was not observed for small ODs. It can also be seen from Fig.~\ref{either-p-or-c} that the peak of the waveform is higher in (a) than in (b). This shows that the effect of the mapping from space domain to time domain is more pronounced when pump laser has a $z$ profile. This is because in the third-order susceptibility (Eq.~\eqref{chi_as_3rd_def}), when the coupling laser has a $z$ profile $f_c(z)$, it appears in the denominator of $\chi^{(3)}$ through $\Omega_c(z)$. Larger values of $f_c(z)$ result in smaller $\chi^{(3)}$, and thus smaller $\psi(\tau)$ (Eq.~\eqref{psi_coupling_z_1}).

%
%--------------------------------------------------
%\begin{figure}[h]
%\centering
%\includegraphics[width=\linewidth]{Fig10.eps}
\begin{figure}
\includegraphics[width=\linewidth]{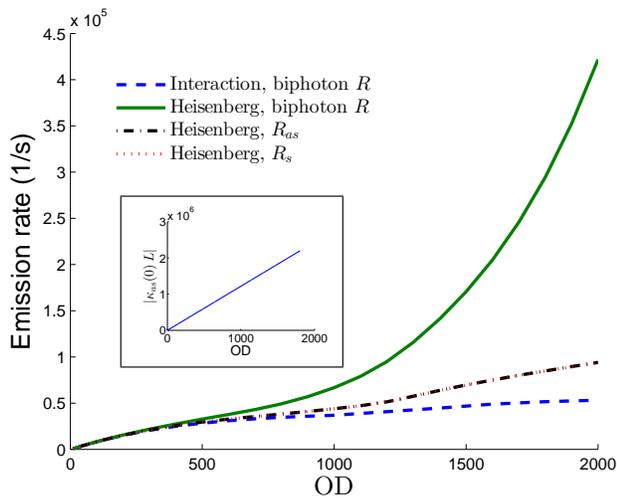}  \\
\caption{(Color online) Comparison of the biphoton generation rate in the interaction (dashed line) and Heisenberg (solid line) pictures as OD increases from 100 to 300 by increasing the atomic density. $x$-axis is the product of $|\kappa_{as}(\varpi =0)\ L|$. This quantity varies linearly with OD (inset figure).}
\label{biphoton_R}
\end{figure}
%%%-------

So far, we have shown theoretically that when there is no $z$-dependence in the atomic density and the driving laser field intensities both model agree well when $q^2 \ll 4\kappa_s\kappa_{as}$, or when the linear response is much larger than the third-order response in the atomic medium. The numerical plots even for non-uniform atomic density and non-uniform driving laser profiles agree as well. Now the question is: what happens when $q^2$ is no longer much larger than $4\kappa_s\kappa_{as}$, and in what region of the parameter space do they differ? To answer this question, we plot photon pair generation rate
as a function of $|\kappa_{as}(\varpi = 0)\, L|$ by varying OD from 100 to 300. We consider homogenous atomic cloud and uniform laser fields. Fig.~\ref{biphoton_R} shows that in the small parametric gain regime where $|\kappa_{as}(\varpi = 0)\, L|$ is small, both models predict the same biphoton rate. However, in the large parametric gain regime where $|\kappa_{as}(\varpi = 0)\, L|$ is large and $q^2 \ll 4\kappa_s\kappa_{as}$ no longer holds, biphoton rate is larger in the Heisenberg picture.

When the third-order response is small, two-photon process dominates. This can be described adequately by the first-order perturbation approximation in the interaction picture. For large $|\kappa_{as}(\varpi = 0)\, L|$, apart from the two-photon process, $n$-photon ($n > 2$) processes are present, this is included in the Heisenberg formalism, but not in the interaction formalism. This is because the first-order perturbation approximation describes only the biphoton process. Therefore, to compare with experimental data in large parametric gain regime, Heisenberg picture should be used. Note that in Figs.~\ref{const_z}, \ref{N-varies-pc-const} and \ref{either-p-or-c}, $|\kappa_{as}(\varpi = 0)\, L|$ = 0.115. Therefore, we can use either model, where biphoton process dominates and the $n$-photon ($n > 2$) processes are negligible. Note also that in the group delay regime, the biphoton joint spectrum is determined by the phase-matching condition. As shown in Fig.~\ref{spectrum}, the phase-matching spectrum function $|F(\varpi)Q(\varpi)|$ is much narrower than the nonlinear gain spectrum $|\kappa_{as}(\varpi)\,L|$. Therefore, we choose $|\kappa_{as}(\varpi=0)\,L|$ as a (dimensionless) parameter to compare the biphoton generation rate in Fig.~\ref{biphoton_R}.

\begin{figure}[h]
\centering
\includegraphics[width=\linewidth]{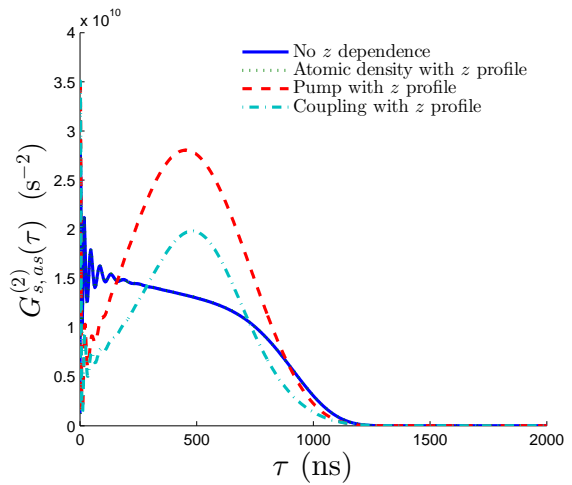}
\caption{(Color online) $G^{(2)}_{s,as}(\tau)$ in the Heisenberg picture for (1) uniform atomic density, pump and coupling laser profiles, (2) atomic density is non-uniform in $z$ direction, with a profile function $f_N(z)=6(z+L/2)(L/2-z)/L^2$ (3) pump laser with a Gaussian profile in $z$ direction, and (4) coupling laser with a Gaussian profile in $z$ direction. The Gaussian profile is $f_{p,c}(z) = 1/0.65  \exp( -z^2/ (L/2)^2)$.}
\label{H-npc}
\end{figure}

In Fig.~\ref{H-npc}, we compare four different situations in the Heisenberg picture: (1) atomic density is uniform, pump and coupling lasers have a uniform $z$ profile, (2) atomic density is non-uniform in $z$ direction, pump and coupling lasers have constant $z$ profiles, (3) the pump laser has a Gaussian profile in $z$ direction, the other quantities are constant in $z$, and (4) the coupling laser has a Gaussian profile in $z$ direction, the other quantities remain constant as $z$ varies. (1) and (2) give the same $G^{(2)}_{s,as}(\tau)$ with a rectangle-like shape. $G^{(2)}_{s,as}(\tau)$ in (3) and (4) is Gaussian-like and the peak is higher for (3).

%--------------------------------------------------------------------------------------------------------
\section{Quantum Waveform Shaping and Engineering} \label{Discussion}
%--------------------------------------------------------------------------------------------------------

\begin{figure}[h]
\begin{center}$
\begin{array}{c}
\hspace{3em} \includegraphics[width=\linewidth]{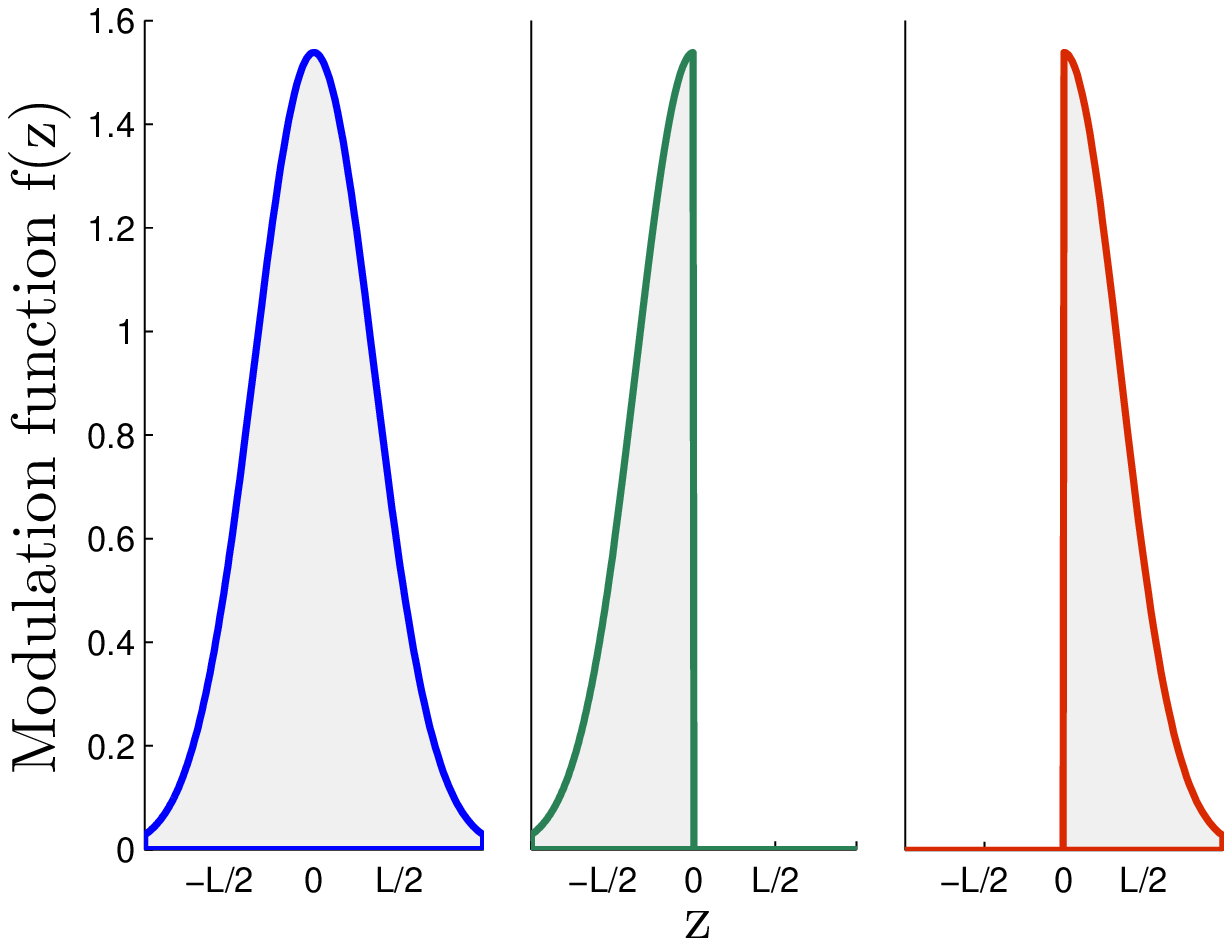}  \\
(a)                                                \\
\includegraphics[width=\linewidth]{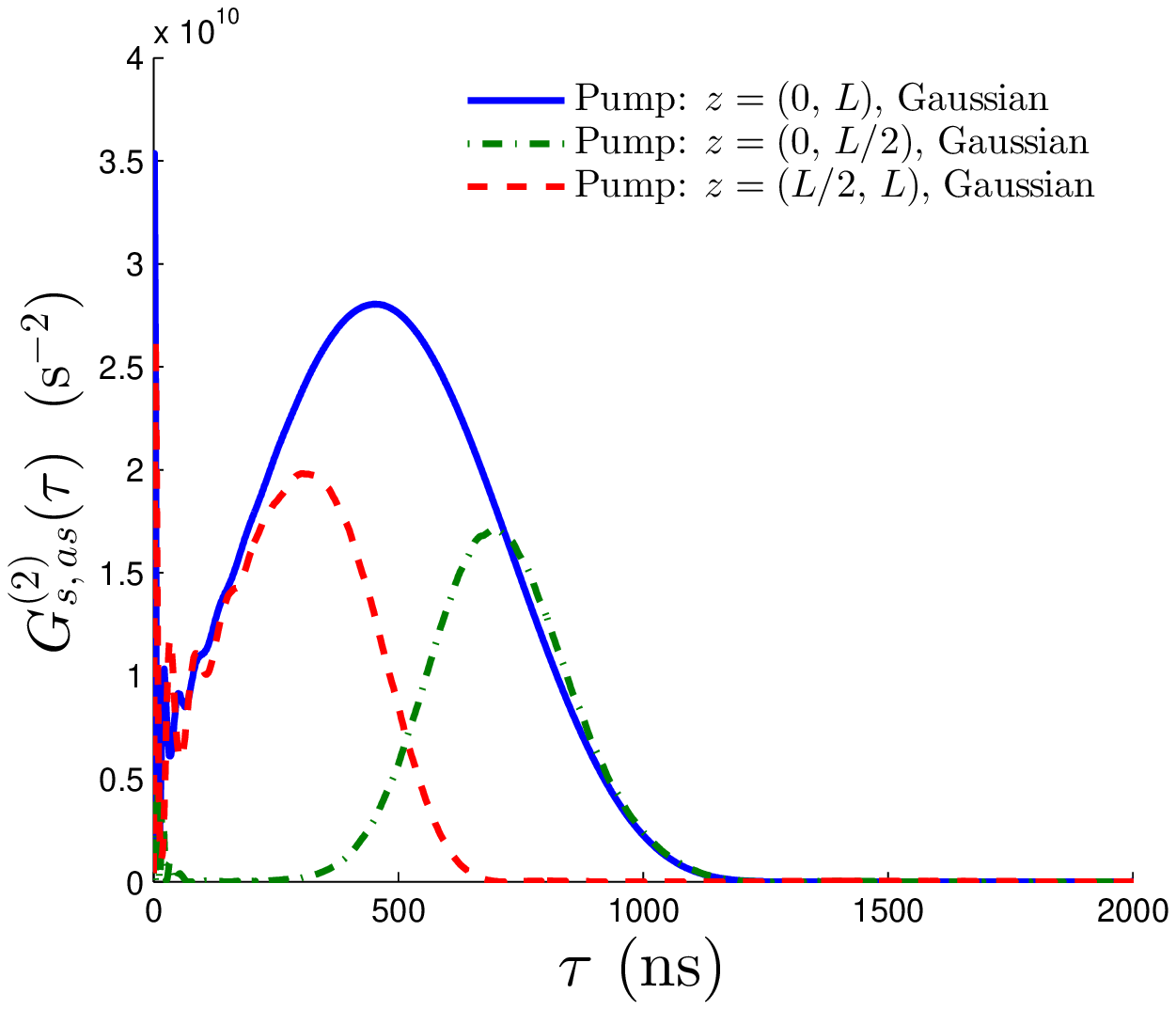} \\
(b)
\end{array}$
\end{center}
\caption{(Color online) (a) The $z$ profile of the pump laser varies from a full Gaussian function $f(z) = 1/0.65 \exp( -z^2/ (L/2)^2)$, to two half-Gaussian functions, $f_l(z)  = 1/0.65 \exp( -z^2/ (L/2)^2)$ for $0<z<L/2$ and zero for $L/2<z<L$,
$f_r(z) = 0$ for $0<z<L/2$ and $f_r(z) = 1/0.65  \exp( -z^2/ (L/2)^2)$ for $L/2<z<L$. (b) $G^{(2)}_{s,as}(\tau)$ corresponding to the pump laser profile as a full-Gaussian $f(z)$, a half blocked Gaussian $f_l(z)$, and another half-blocked Gaussian $f_r(z)$.  }
\label{pump-Gaussian}
\end{figure}
\begin{figure}[h]
\begin{center}$
\begin{array}{c}
  \hspace{3em}\includegraphics[width=\linewidth]{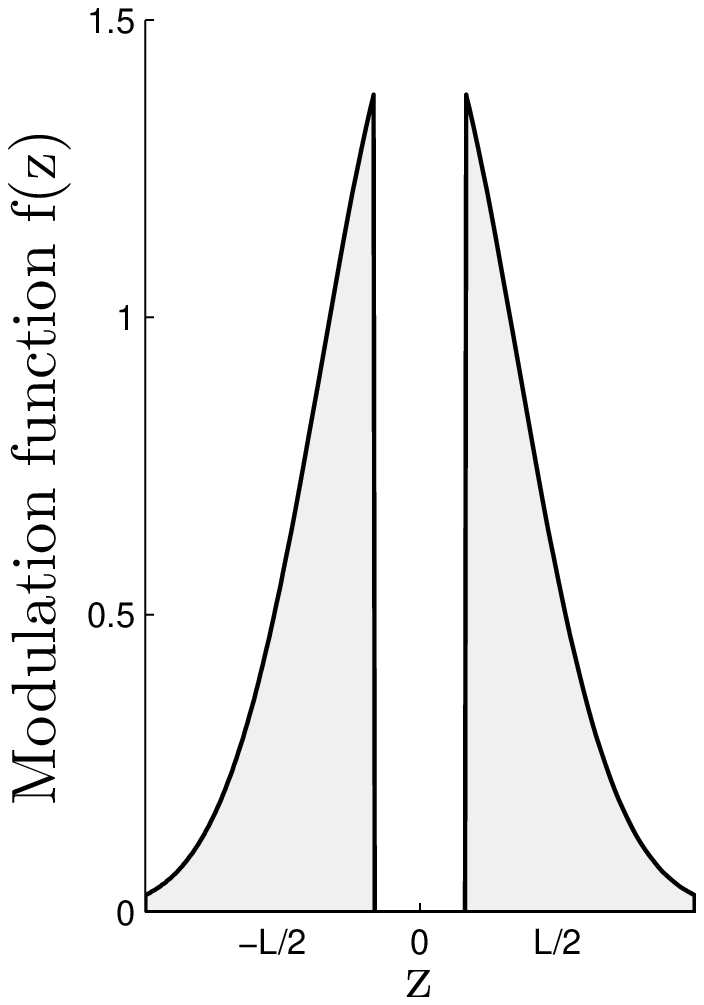} \\
(a) \\
 \includegraphics[width=\linewidth]{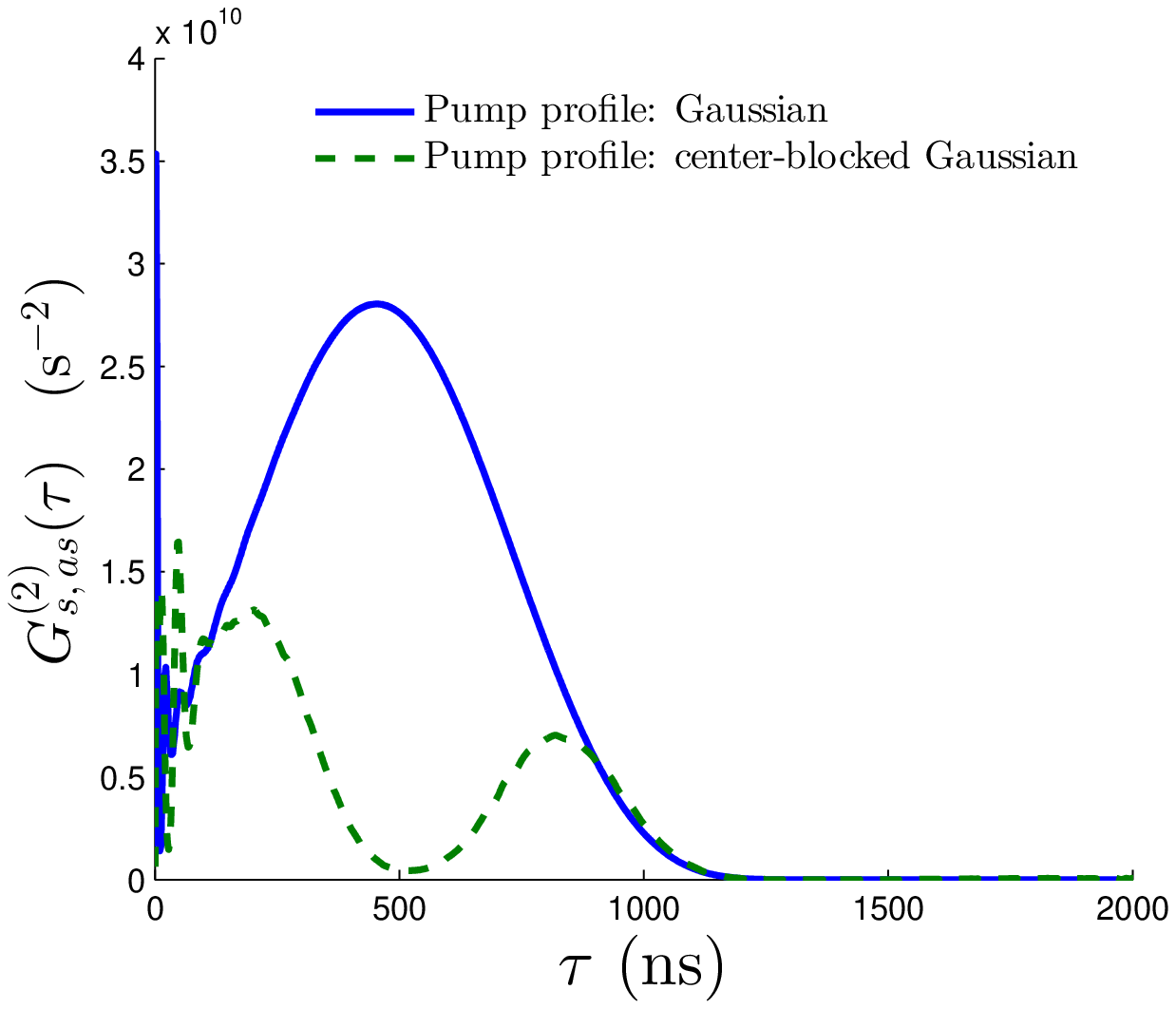} \\
(b)
\end{array}$
\end{center}
\caption{(Color online) (a) The $z$ profile of the pump laser varies from a center-blocked Gaussian function $f_p(z) = 1/0.65\ \exp( -z^2/ (L/2)^2)$, for $L/3 < z < 2L/3$ and zero otherwise. (b) The corresponding $G^{(2)}_{s,as}(\tau)$ for the blocked and full Gaussian profiles. }
\label{pump-center-blocked}
\end{figure}

In this section, we explore possibilities of manipulating the biphoton temporal waveform by tailoring the pump laser spatial profiles. We keep the atomic density and the coupling laser profile uniform in space. Figure \ref{pump-Gaussian}(a) shows a pump laser with three $z$ profiles: (1) a full Gaussian function $f_p(z) = 1/0.65 \exp( -z^2/ (L/2)^2)$, (2) and (3) half-Gaussian profile, with either the left or the right side of the beam is covered from the center of the Gaussian curve. This means that only half of the atomic cloud is exposed to the pump laser. The corresponding $G^{(2)}_{s,as}(\tau)$ is plotted in Fig.~\ref{pump-Gaussian}(b). It is expected that when half of the laser beam is covered, $G^{(2)}_{s,as}(\tau)$ should be lower than that when the full beam is present. However, it is interesting that when the MOT that lies in $(0,L/2)$ is exposed to the half-Gaussian beam, the nonzero part of $G^{(2)}_{s,as}(\tau)$ in space domain is shifted to the larger delay part of time domain, and vice verse. This is as explained in section \ref{section:pump_z}.

Next, we block the center part of the Gaussian beam, i.e., the part of the atomic cloud that lies in $(L/3, 2L/3)$ are not subjected to the pump laser beam (Fig.~\ref{pump-center-blocked}(a)). $G^{(2)}_{s,as}(\tau)$ is shown in Fig.~\ref{pump-center-blocked}(b). There are two bumps present corresponding to the two parts of the pump laser profile, as well as the biphoton optical precursor. For comparison, we also show the case when the pump laser has a full Gaussian profile.

Experimentally, if we can control the profile of the driving laser fields spatially, we can produce interesting biphoton waveforms temporally. These interesting waveforms might open up more applications in quantum information technology. For example, information in spatial pattern of the driving fields can be coded onto the temporal pattern of the biphoton waveform. On the other hand, a time-domain biphoton waveform allows us to deduce the information on the space-domain profile of the driving lasers, so a desired time-domain pattern can be obtained with corresponding modulations on the laser profiles in space domain.

%------------------------------------------------------------------------------
\section{Conclusions} \label{conclusion}
%------------------------------------------------------------------------------

We have generalized and compared the theoretical modeling of the biphoton generation through the SFWM process.  We showed that both approaches in the interaction and Heisenberg pictures agree well in low parametric gain regime. Moreover, when the pump and coupling lasers have nonuniform profile in the atomic medium, the second-order correlation function of Stokes--anti-Stokes is no longer rectangle-like with a modified exponential tail, but Gaussian-like with a peak in the middle. This is confirmed by recent experimental data. We also predicted that one can control the shape of the time-domain biphoton waveform by tailoring the space-domain profile of the pump and coupling lasers, especially the pump profile as it dominates the effect from space-to-time mapping.
%------------------------------------------------------------------------------------------------------------------------

\begin{acknowledgments}
The work was supported by the Hong Kong Research Grants Council (Project No. 601113).
\end{acknowledgments}

%----------------------------------------------------------------------------------------------------------------------
\appendix
\numberwithin{equation}{section}

%----------------------------------------------------------------------------------------------------
\section{Solution to the coupled ODE in the Heisenberg-Langevin formalism}

As the susceptibilities are functions of $\varpi$, $\alpha_{as}$,  $g_s$, $\kappa_{as}$ and $\kappa_s$ are also functions of $\varpi$.

The general solution to \eqref{ode_omega} at $z=l$ can be written as
\begin{eqnarray} \label{Solution1}
& \left( \begin{array}{c} \hat{a}_{as}(l,\varpi) \\ \hat{a}_{s}^\dag (l,-\varpi)  \end{array} \right)
= \begin{pmatrix}
    A_1(\varpi) & B_1(\varpi) \\
	  C_1(\varpi) & D_1(\varpi)
	\end{pmatrix}
		\left( \begin{array}{c} \hat{a}_{as}(-l,\varpi) \\ \hat{a}_{s}^\dag (-l,-\varpi)  \end{array} \right)
\nonumber\\
&
 + \sum_{\alpha_i}\int_{-l}^l dz\ e^{M(z-l)} \begin{pmatrix} \beta_{\alpha_i}^{as}\\ \beta_{\alpha_i}^s\end{pmatrix}
\hat{f}_{\alpha_i^\dag}.
\end{eqnarray}
where the transform matrix can be obtained by numerically solving the coupled equation by setting the Langevin forces to zero. $M$ is given by
\begin{equation}\label{MatrixM}
M= \begin{pmatrix}
     \alpha_{as}(z,\varpi)-i\frac{\Delta k_0}{2} & -\kappa_{as}(z,\varpi) \\
		 -\kappa_{s}(z,\varpi) & g_{s}(z,\varpi)+i\frac{\Delta k_0}{2}
	\end{pmatrix}.
\end{equation}
%%%%%%%%%%%%%%%%%%%%%%%%%%%%%%%%%%%%%%%%%%%%%%%%%%%%%%%%%%%%%%%%%%%%%%%%%%%%%%%

Equation \eqref{Solution1} can be rewritten as Eq.~\ref{0L}, where
\begin{equation} \label{ABCD}
\begin{split}
	& A(\varpi)  = A_1(\varpi) - \frac{B_1(\varpi) C_1(\varpi)}{D_1(\varpi)} \\
	& B(\varpi)  = \frac{B_1(\varpi)}{D_1(\varpi)} \\
	& C(\varpi)  = - \frac{C_1(\varpi)}{D_1(\varpi)} \\
	& D(\varpi)  = \frac{1}{D_1(\varpi)},
\end{split}
\end{equation}
and
\begin{equation}
\begin{pmatrix} P_{\alpha_i} \\ Q_{\alpha_i} \end{pmatrix}
=
\begin{pmatrix} 1 & -\frac{B_1(\varpi)}{D_1(\varpi)} \\
                0 & -\frac{1}{D_1(\varpi)}
\end{pmatrix}
e^{M(z-L)}
\begin{pmatrix} \beta_{\alpha_i}^{as} \\ \beta_{\alpha_i}^s \end{pmatrix} .
\end{equation}

The two-photon correlation Glauber function is given by
\begin{equation}\label{G2}
\begin{split}
& G^{(2)}_{s,as}(t,t+\tau) =
   \langle \hat{a}_{as}^\dag(l,t+\tau) \hat{a}_s^\dag(-l,t) \hat{a}_s(-l,t) \hat{a}_{as}(l,t+\tau) \rangle \\
&	=  \frac{1}{(2\pi)^2}\int d\varpi_1 d\varpi_2 d\varpi_3 d\varpi_4 e^{i\varpi_1(t+\tau) - i\varpi_2 t + i\varpi_3 t - i\varpi_4(t+\tau)}  \\
&\ \ \ 	\times \langle \hat{a}_{as}^\dag(l,\varpi_1) \hat{a}_{s}^\dag(-l,-\varpi_2) \hat{a}_{s}(-l,-\varpi_3) \hat{a}_{as}(l,\varpi_4) \rangle.
\end{split}
\end{equation}

From Eq.~\eqref{0L} and the boundary condition (Eq.~\eqref{Boundarycondition2}), assuming the starting time $t=0$, the Glauber function is then
\begin{equation}\label{G2_2}
\begin{split}
& G^{(2)}_{s,as}(\tau) =	
\\
& \bigg| \int \frac{d\varpi}{2\pi} e^{-i\varpi\tau}  \Big[B(\varpi) D^*(\varpi)
              + \sum\limits_{\alpha_i,\alpha_j}\int\limits_{-l}^l dz
									Q^*_{\alpha_i}D_{\alpha_i,\alpha_j^\dag}P_{\alpha_j} \Big] \bigg|^2
\\
& \  + R_{as}R_s.	
\end{split}						
\end{equation}

The second term in \eqref{G2_2} is the product of Stokes and anti-Stokes generation rates, $R_s$ and $R_{as}$.
This term describes a uniform background.

The Langevin noise coefficients are given by
\begin{equation}
\beta_{21}^{as} = -\frac{\sqrt{2}\,\Omega_c(z)\sqrt{N(z)\sigma_{13}\gamma_{13}}}{G(\varpi)},
\end{equation}
\begin{equation}
\beta_{24}^{as} = \left(\frac{\Omega_p(z)}{\Delta_p}\right)
                  \frac{\Omega_c(z)\sqrt{N(z)\sigma_{13}\gamma_{13}}}{\sqrt{2}\,G(\varpi)},
\end{equation}
\begin{equation}
\beta_{31}^{as} = \frac{2\sqrt{2}\,(\varpi+i\gamma_{12})\sqrt{N(z)\sigma_{13}\gamma_{13}}}{G(\varpi)},
\end{equation}
\begin{equation}
\beta_{34}^{as} = -\left(\frac{\Omega_p(z)}{\Delta_p}\right)
                   \frac{\sqrt{2}\,(\varpi+i\gamma_{12})\sqrt{N(z)\sigma_{13}\gamma_{13}}}{G(\varpi)},
\end{equation}
\begin{equation}
\beta_{21}^{s} = -\left(\frac{\Omega_p(z)}{\Delta_p}\right)
                  \frac{\sqrt{2}\,(\varpi + i\gamma_{13})\sqrt{N(z)\sigma_{24}\gamma_{24}}}{G(\varpi)},
\end{equation}
\begin{equation}
\beta_{24}^{s} = \frac{\sqrt{N(z)\sigma_{24}\gamma_{24}}}{\sqrt{2}\,\Delta_p},
\end{equation}
\begin{equation}
\beta_{31}^{s} = \left(\frac{\Omega_p(z)}{\Delta_p}\right)
                  \frac{\Omega_c(z)\sqrt{N(z)\sigma_{24}\gamma_{24}}}{\sqrt{2}\,G(\varpi)},
\end{equation}
\begin{equation}
\beta_{34}^{s} = \frac{\Omega_c(z)\sqrt{N(z)\sigma_{24}\gamma_{24}}}{2\sqrt{2}\,\Delta_p^2}.
\end{equation}
Here $G(\varpi) \equiv |\Omega_c(z)|^2 - 4(\varpi + i\gamma_{12})(\varpi + i\gamma_{13})$, $\sigma_{ij}$ is the absorption cross section for $|j\rangle \rightarrow |i\rangle$ transition. $\gamma_{ij}$ is the dephasing rate between $|i\rangle$ and $|j\rangle$.

The diffusion coefficients are given by
\begin{widetext}
\begin{equation}\label{D_ai_ajc}
D_{\alpha_i,\alpha_j^\dag} = \begin{pmatrix}
									2\langle \tilde{\sigma}_{22}\rangle \gamma_{12}
									+ 2\langle \tilde{\sigma}_{33}\rangle \gamma_{23}
									+ 2\langle \tilde{\sigma}_{44}\rangle\gamma_{24} & 0 & \langle \tilde{\sigma}_{23}\rangle\gamma_{12} & 0
									\\
									0 & 2\langle \tilde{\sigma}_{22}\rangle (\gamma_{14} + \gamma_{24})
									    +  2\langle \tilde{\sigma}_{33}\rangle \gamma_{23}
											+ 2\langle \tilde{\sigma}_{44}\rangle \gamma_{24}
										& 0 & 2\langle \tilde{\sigma}_{23}\rangle (\gamma_{14}+\gamma_{24})
									\\
									\langle \tilde{\sigma}_{32} \rangle \gamma_{12} & 0 & 0 & 0
									\\
									0 & 2\langle \tilde{\sigma}_{32}\rangle (\gamma_{14}+\gamma_{24})
									  & 0 & 2\langle \tilde{\sigma}_{33} \rangle (\gamma_{14}+\gamma_{24})
	\end{pmatrix},
\end{equation}
\end{widetext}
\begin{widetext}
\begin{equation}\label{D_aic_aj}
D_{\alpha_i^\dag,\alpha_j} = \begin{pmatrix}
									2\langle \tilde{\sigma}_{11}\rangle \gamma_{12}
									+ 2\langle \tilde{\sigma}_{33}\rangle  \gamma_{13}
									+ 2\langle \tilde{\sigma}_{44}\rangle\gamma_{14}
									& \langle \tilde{\sigma}_{14} \rangle \gamma_{12} & 0 & 0
									\\
									\langle \tilde{\sigma}_{41} \rangle \gamma_{12} & 0 & 0 & 0								
									\\
									0 & 0 &
									2\langle \tilde{\sigma}_{11}\rangle (\gamma_{13}+\gamma_{23})
									+ 2\langle \sigma_{33}\rangle \gamma_{13}
									+ 2\langle \tilde{\sigma}_{44}\rangle\gamma_{14}
									& 2\langle \tilde{\sigma}_{14} \rangle (\gamma_{13} + \gamma_{23})									
									\\
									0 & 0 & 2\langle \tilde{\sigma}_{41} \rangle (\gamma_{13} + \gamma_{23})
									  & 2\langle \tilde{\sigma}_{44} \rangle (\gamma_{13} + \gamma_{23})
	\end{pmatrix},
\end{equation}
\end{widetext}
with $\alpha_i$ denoting ${21,24,31,34}$ and $\alpha_i^\dag$ denoting ${12,42,13,43}$. The expectation values of atomic operators in Eqs.~\ref{D_ai_ajc} and \ref{D_aic_aj} are given by
\begin{equation}
\langle \tilde{\sigma}_{11} \rangle = \frac{2\gamma_{13}|\Omega_c(z)|^2 \left[4(\gamma_{14}+\gamma_{24})^2 + 4 \Delta_p^2 + |\Omega_p(z)|^2\right]}{T},
\end{equation}
\begin{equation}
\langle \tilde{\sigma}_{22} \rangle = \frac{2\gamma_{24}\left[4(\gamma_{13}+\gamma_{23})^2 + |\Omega_c(z)|^2\right]|\Omega_p(z)|^2}{T},
\end{equation}
\begin{equation}
\langle \tilde{\sigma}_{33} \rangle = \frac{2\gamma_{24}|\Omega_c(z)\Omega_p(z)|^2}{T},
\end{equation}
\begin{equation}
\langle \tilde{\sigma}_{44} \rangle = \frac{2\gamma_{13}|\Omega_c(z)\Omega_p(z)|^2}{T},
\end{equation}
\begin{equation}
\langle\tilde{\sigma}_{14} \rangle = -\frac{4\gamma_{13}\left[\Delta_p - i(\gamma_{14}+\gamma_{24})\right]|\Omega_c(z)|^2\Omega_p(z)}{T},
\end{equation}
\begin{equation}
\langle \tilde{\sigma}_{23} \rangle = \frac{i4(\gamma_{13}+\gamma_{23})\gamma_{24}\Omega_c(z)|\Omega_p(z)|^2}{T},
\end{equation}
where
\begin{widetext}
\begin{equation}
T = 4\gamma_{13}\left[2(\gamma_{14}+\gamma_{24})^2 + 2\Delta_p^2 + \Omega_p(z)|^2\right]|\Omega_c(z)|^2
+ 4\gamma_{24}\left[2(\gamma_{13}+\gamma_{23})^2 + |\Omega_c(z)|^2\right]|\Omega_p(z)|^2.
\end{equation}
\end{widetext}
The diffusion matrix in Eq.~\ref{D_ai_ajc} contains only excited state populations, it will be thus very small when the pump is weak and far-detuned. In this case, its contribution to the Glauber function will be small.

%------------------------------------------------------------


\begin{thebibliography}{[1]}

\bibitem{BraunsteinRMP2005} S.L. Braunstein, and P. van Loock, ``Quantum information with continuous variables,'' Rev. Mod. Phys. \textbf{77}, 513 (2005).

\bibitem{GisinRMP2002} N. Gisin, G. Ribordy, W. Tittel, and H. Zbinden, ``Quantum cryptography,'' Rev. Mod. Phys. \textbf{74}, 145--195 (2002).

\bibitem{OuPRL1999} Z.Y. Ou, and Y.J. Lu, ``Cavity Enhanced Spontaneous Parametric Down-Conversion for the Prolongation of Correlation Time between Conjugate Photons,'' Phys. Rev. Lett. \textbf{83}, 2556--2559 (1999).

\bibitem{KuklewiczPRL2006} C.E. Kuklewicz, F.N. Wong, and J.H. Shapiro, ``Time-Bin-Modulated Biphotons from Cavity-Enhanced Down-Conversion,'' Phys. Rev. Lett. \textbf{97}, 223601 (2006).

\bibitem{PanPRL2008} X.-H. Bao, Y. Qian, J. Yang, H. Zhang, Z.-B. Chen, T. Yang, and J.-W. Pan, ``Generation of Narrow-Band Polarization-Entangled Photon Pairs for Atomic Quantum Memories,'' Phys. Rev. Lett. \textbf{101}, 190501 (2008).

\bibitem{ChuuPRA2011} C.-S. Chuu and S. E. Harris, ``Ultrabright backward-wave biphoton source,'' Phys. Rev. A \textbf{83}, 061803(R) (2011).

\bibitem{ChuuAPL2012} C.-S. Chuu, G. Y. Yin, and S. E. Harris, ``A miniature ultrabright source of temporally long, narrowband biphotons,'' Appl. Phys. Lett. \textbf{101}, 051108 (2012).

\bibitem{BalicPRL2005} V. Balic, D. Braje, P. Kolchin, G. Y. Yin, and S. E. Harris, ``Generation of Paired Photons with Controllable Waveforms,'' Phys. Rev. Lett. \textbf{94}, 183601 (2005).

\bibitem{DuPRL2007} S. Du, J. Wen, M. H. Rubin, and G.Y. Yin, ``Four-Wave Mixing and Biphoton Generation in a Two-Level System,'' Phys. Rev. Lett. \textbf{98}, 053601 (2007).

\bibitem{DuPRL2008} S. Du, P. Kolchin, C. Belthangady, G.Y. Yin, and S.E. Harris, ``Subnatural Linewidth Biphotons with Controllable Temporal Length,'' Phys. Rev. Lett. \textbf{100}, 183603 (2008).

\bibitem{SrivathsanPRL2013} B. Srivathsan, G.K. Gulati, B. Chng, G. Maslennikov, D. Matsukevich, and C. Kurtsiefer, ``Narrow Band Source of Transform-Limited Photon Pairs via Four-Wave Mixing in a Cold Atomic Ensemble,'' Phys. Rev. Lett. \textbf{111}, 123602 (2013).

\bibitem{ZhaoOPTICA2014} L. Zhao, X. Xian, C. Liu, Y. Sun, M. M. T. Loy, and S. Du, ``Photon pairs with coherence time exceeding 1 $\mu$s,'' Optica \textbf{1}, 84 (2014).

\bibitem{SinglePhotonEOM} P. Kolchin, C. Belthangady, S. Du, G. Y. Yin, and S. E. Harris, ``Electro-Optic Modulation of Single Photons,'' Phys. Rev. Lett. \textbf{101}, 103601 (2008).

\bibitem{SinglePhotonPrecursor} S. Zhang, J. F. Chen, C. Liu, M. M. T. Loy, G. K. L. Wong, and S. Du, ``Optical Precursor of a Single Photon,'' Phys. Rev. Lett. \textbf{106}, 243602 (2011).

\bibitem{ZhouOE2012} S. Zhou, S. Zhang, C. Liu, J. F. Chen, J. Wen, M. M. T. Loy, G. K. L. Wong, and S. Du, ``Optimal storage and retrieval of single-photon waveforms,'' Opt. Express \textbf{20}, 24124-24231 (2012).

\bibitem{ZhangPRL2012} S. Zhang, C. Liu, S. Zhou, C.-S. Chuu, M.M.T. Loy, and S. Du, ``Coherent Control of Single-Photon Absorption and Reemission in a Two-Level Atomic Ensemble,'' Phys. Rev. Lett. \textbf{109}, 263601 (2012).

\bibitem{LiuOE2013} C. Liu, S. Zhang, L. Zhao, P. Chen, C.-H. Fung, H. Chau, M. Loy, and S. Du, ``Differential-phase-shift quantum key distribution using heralded narrow-band single photons,'' Opt. Express \textbf{21}, 9505-9513 (2013).

\bibitem{ValenciaPRL2007} A. Valencia, A Cer, X. Shi, G. Molina-Terriza, and J. Torres, ``Shaping the Waveform of Entangled Photons,'' Phys. Rev. Lett. \textbf{99}, 243601 (2007).

\bibitem{DuPRA2009} S. Du, J. Wen, and C. Belthangady, ``Temporally shaping biphoton wave packets with periodically modulated driving fields,'' Phys. Rev. A \textbf{79}, 043811 (2009).

\bibitem{ChenPRL2010} J.F. Chen, S. Zhang, H. Yan, M. M. Loy, G. K. Wong, and S. Du, ``Shaping Biphoton Temporal Waveforms with Modulated Classical Fields,'' Phys. Rev. Lett. \textbf{104}, 183604 (2010).

%Interaction picture reference
\bibitem{Wen2006}
J.M. Wen and M.H. Rubin, ``Transverse effects in paired-photon generation via an electromagnetically induced transparency medium. I. Perturbation theory,'' Phys. Rev. A \textbf{74}, 023808 (2006).

\bibitem{Wen2007_1}
J.M. Wen, S. Du, and M.H. Rubin, ``Biphoton generation in a two-level atomic ensemble,'' Phys. Rev. A \textbf{75}, 033809 (2007).

\bibitem{Wen2007_2}
J.M. Wen, S. Du, and M.H. Rubin, ``Spontaneous parametric down-conversion in a three-level system,'' Phys. Rev. A \textbf{76}, 013825 (2007).

\bibitem{Wen2008}
J.M. Wen, S. Du, Y. Zhang, M. Xiao, and M.H. Rubin, ``Nonclassical light generation via a four-level inverted-Y system,'' Phys. Rev. A \textbf{77}, 033816 (2008).

\bibitem{Du}
S. Du, J.M. Wen, M.H. Rubin, ``Narrowband biphoton generation near atomic resonance,'' J. Opt. Soc. Am. B25, C98--C108 (2008).

%Heisenberg picture reference
\bibitem{Kolchin}
P. Kolchin, ``Electromagnetically-induced-transparency-based paired photon generation,'' Phys. Rev. A \textbf{75}, 033814 (2007).

\bibitem{Ooi}
C.H.R. Ooi, Q. Sun, M.S. Zubairy and M.O, Scully, ``Correlation of photon pairs from the double Raman amplifier: Generalized analytical quantum Langevin theory,'' Phys. Rev. A \textbf{75}, 013820 (2007).

\bibitem{WenRubin}
J.M. Wen and M.H. Rubin, ``Transverse effects in paired-photon generation via an electromagnetically induced transparency medium. II. Beyond perturbation theory,'' Phys. Rev. A \textbf{74}, 023809 (2006).

\bibitem{Braje}
D.A. Braje, V. Balic, S. Goda, G.Y. Yin and S.E. Harris, ``Frequency Mixing Using Electromagnetically Induced Transparency in Cold Atoms,'' Phys. Rev. Lett. \textbf{93}, 183601 (2004).


%EIT reference
\bibitem{EIT} S. E. Harris, ``Electromagnetically induced transparency,'' Phys. Today \textbf{50}, 36-40 (1997).

\bibitem{EIT02} M. Fleischhauer, A. Imamoglu, and J. P. Manarangos, ``Electromagnetically induced transparency: Optics in coherent media,'' Rev. Mod. Phys. \textbf{77}, 633--673 (2005).

\bibitem{Rubin} M. H. Rubin, D. N. Klyshko, Y. H. Shih, and A. V. Sergienko, ``Theory of two-photon entanglement in type-II optical parametric down-conversion,'' Phys. Rev. A. \textbf{50}, 5122--5133 (1994).

%reference for cauchy-schwarz inequality
\bibitem{Clauser1974} J. F. Clauser, ``Experimental distinction between the quantum and classical field-theoretic predictions for the photoelectric effect,'' Phys. Rev. D \textbf{9}, 853--860 (1974).



\end{thebibliography}
\end{document}